\documentclass[british,aps,manuscript]{revtex4-1}
\usepackage[T1]{fontenc}
\usepackage[latin9]{inputenc}
\usepackage{geometry}
\usepackage{amsmath}
\usepackage{amssymb}
\usepackage{graphicx}
\usepackage{esint}

\makeatletter
 
 \@ifundefined{textcolor}{}
 {%
   \definecolor{BLACK}{gray}{0}
   \definecolor{WHITE}{gray}{1}
   \definecolor{RED}{rgb}{1,0,0}
   \definecolor{GREEN}{rgb}{0,1,0}
   \definecolor{BLUE}{rgb}{0,0,1}
   \definecolor{CYAN}{cmyk}{1,0,0,0}
   \definecolor{MAGENTA}{cmyk}{0,1,0,0}
   \definecolor{YELLOW}{cmyk}{0,0,1,0}
 }
\numberwithin{equation}{section}
\DeclareMathAlphabet\mathbfcal{OMS}{cmsy}{b}{n}

\makeatother

\usepackage{babel}
\begin{document}

\title{$c-$number Quantum Generalised Langevin Equation for an open system}

\author{L. Kantorovich\thanks{lev.kantorovitch@kcl.ac.uk}, H. Ness, L. Stella$^{\dagger}$
and C. Lorenz}

\address{Physics Department, King's College London, Strand, London, WC2R 2LS,
United Kingdom\\$^{\dagger}$Atomistic Simulation Centre, School
of Mathematics and Physics, Queen\textquoteright s University Belfast,
University Road, Belfast BT7 1NN, Northern Ireland, UK}
\begin{abstract}
We derive a $c-$number Generalised Langevin Equation (GLE) describing
the evolution of the expectation values $\left\langle x_{i}\right\rangle _{t}$
of the atomic position operators $x_{i}$ of an open system. The latter
is coupled linearly to a harmonic bath kept at a fixed temperature.
The equations of motion contain a non-Markovian friction term with
the classical kernel {[}L. Kantorovich - PRB 78, 094304 (2008){]}
and a zero mean \emph{non-Gaussian} random force with correlation
functions that depend on the initial preparation of the open system.
We used a density operator formalism without assuming that initially
the combined system was decoupled. The only approximation made in
deriving quantum GLE consists in assuming that the Hamiltonian of
the open system at time $t$ can be expanded up to the second order
with respect to operators of atomic displacements $u_{i}=x_{i}-\left\langle x_{i}\right\rangle _{t}$
in the open system around their exact atomic positions $\left\langle x_{i}\right\rangle _{t}$
(the ``harmonisation'' approximation). The noise is introduced to
ensure that sampling many quantum GLE trajectories yields exactly
the average one. An explicit expression for the pair correlation function
of the noise, consistent with the classical limit, is also proposed.
Unlike the usually considered quantum operator GLE, the proposed $c-$number
quantum GLE can be used in direct molecular dynamic simulations of
open systems under general equilibrium or non-equilibrium conditions. 
\end{abstract}
\maketitle

\section{Introduction}

In numerous applications in quantum physics and chemistry \cite{Petuccione-Open-Systems-2007,Weiss-2012},
phenomena of interest are related to an atomistic dynamics of a finite
fragment of an extended system. The fragment cannot be treated as
isolated as it interacts and exchanges energy with the rest of the
system serving as a heat bath. As a complete description of the whole
system might be difficult or impossible, one has to look for approaches
which pay specific attention to the fragment (an open system), while
still retaining the existence of the heat bath. This problem lies
within the realm of \emph{open quantum dissipative systems} \cite{Petuccione-Open-Systems-2007,Weiss-2012}.

In the case of classical systems, molecular dynamics (MD) simulations
have proven to be a powerful, yet simple, tool for studying their
non-equilibrium properties including tribology \cite{Szlufarska2007,Barry2009,Lorenz2010},
energy dissipation \cite{Trevethan2004}, crack propagation \cite{Kermode2008},
heat transport \cite{Mazyar2006,Hu2008,Dhar-Roy-JSP-2008,Hu2009,Guo2010,Hu2011,Manikandan2011}
and irradiation \cite{Hsu2007}. An appropriate theoretical approach
for considering dynamics of open \emph{classical} systems, based on
calculating trajectories of atoms of the open system and accounting
for dissipation effects with its environment(s), is provided by the
Generalised Langevin Equation (GLE) \cite{Zwanzig2001}. Assuming
a rather general Hamiltonian of the open system and linear coupling
to its harmonic heat bath, one arrives at its non-Markovian classical
dynamics with multivariate Gaussian distributed random forces and
the memory kernel that is proportional to the random force autocorrelation
function (the second fluctuation-dissipation theorem) \cite{my-SBC-1}.
Although the GLE has been around for a while (see \cite{my-SBC-1,Lorenzo-GLE-2014,Herve-GLE-2015}
and references therein), its application to realistic systems has
only recently become realised when a powerful implementation of this
method has been proposed \cite{Lorenzo-GLE-2014,Herve-GLE-2015}.
It solved two main obstacles standing in the way of efficient numerical
simulations: its non-Markovian character and the coloured noise. This
implementation is also straightforwardly generalised for heat transport
simulations which require more than one heat bath \cite{Herve-GLE-2016}.

Although classical MD simulations for open systems can be easily justified
via GLE, a natural question arises of whether something analogous
to GLE can also be formulated in the quantum realm. By that we mean
equations of motion for the \emph{expectation values }of positions
of atoms in the open system, $\left\langle x_{i}\right\rangle _{t}=\mbox{Tr}\left(\rho(t)x_{i}\right)$
(where $\rho(t)$ is the exact density matrix at time $t$ of the
whole combined system, and $x_{i}$ the operator of the coordinate
of atom $i$); the equations are expected to contain a \emph{non-operator}
(or $c-$number) stochastic force with certain statistical properties.
At high enough temperatures (or in the $\hbar\rightarrow0$ limit)
the $c-$number quantum GLE must coincide with the classical one for
the same Hamiltonian \cite{my-SBC-1}. Using such a tool, one will
be able to study, via MD-like approaches, dynamical phenomena of a
wide range of systems e.g. in quantum optics, condensed matter and
chemical physics and nanotechnology, accounting for the full quantum
nature of both the bath and the open system.

First attempts to develop a quantum analogue of the classical GLE
based on equations of motion for Heisenberg positions and momenta
operators of atoms of an open system were done by Ford, Kac and Mazur
(FKM) \cite{FKM-J.Math.Ph.-1965}. This method has been further developed
by other authors \cite{Benguria-Kac-1981,Lindenberg-West-1984,Cortes-West-Lindenberg-JCP-1985,Ford-Kac-JST-1987,Ford-Lewis-PRB-1988,Gardiner-1988,van_Kampen-1997,McDowell-JCP-2000,Nieuwenhuizen-PRE-2002,Yan-Xu-review-2005,Gardiner-book_noise-2010,Ferialdi-Durr-PRA-2015}
and then applied, in particular, to heat transport \cite{Segal2002,Segal-Nitzan-Hanggi-2003,Dhar-Roy-JSP-2008}.
In this method, GLE-like non-Markovian equations for operators of
the open system are obtained assuming a linear coupling to the harmonic
bath. The bath enters these equations via an operator which has a
meaning of a force, the latter contains a linear combination of initial
bath operators. Assuming that initially the bath was at equilibrium
at a certain temperature, and hence the reduced density matrix of
the bath (the density matrix of the whole system in which open system
degrees of freedom are traced out) is canonical at a certain temperature,
one can show that the statistically averaged operator of the force
is zero and its correlation function is essentially given by the well-known
displacement-displacement correlation function calculated in the harmonic
approximation. Although this method is exact within the adopted Hamiltonian,
analytical solutions can only be obtained in simple cases \cite{Gardiner-book_noise-2010}.
It is essential, that even though the equations themselves are written
only for operators of the open system, these operators are still defined
in the Hilbert space of the whole system (open system + bath). For
harmonic systems, this approach for heat transport has been shown
\cite{Dhar-Roy-JSP-2008} to be fully equivalent to the method based
on non-equilibrium Green's functions \cite{Ozpineci-Ciraci-2001,Wang-Wang-Zeng-2006,Galperin-Nitzan-Ratner-2007}. 

Note that equations for quantum operators of an open system with linearly
coupled harmonic bath can also be derived using path-integral techniques
by integrating out the bath variables \cite{Feynman-Vernon-1963,Sebastian-CPL-1981,Eckern-JSP-1990,Kleinert-Shabanov-PLA-1995,Weiss-2012}.
This method is however based on the so-called partitioned assumption
(initiated by Feynman and Vernon \cite{Feynman-Vernon-1963}) that
the initial density matrix is a direct product of independent density
matrices of the open system and bath (the Born approximation \cite{Petuccione-Open-Systems-2007})
.

Several attempts have also been made to obtain truly $c-$number quantum
GLE (cQGLE). In a hybrid approach \cite{Wang-PRL-2007,Dammak-PRL-2009}
the authors suggested simply to replace, without proper justification,
Heisenberg equations of motion for operators of positions of atoms
of an open system by their classical analogues keeping, at the same
time, the fully quantum expression for the random force autocorrelation
function (note that in this approach the random force is not an operator).

A more elaborate approach based on a somewhat artificial distribution
function for the bath has also been proposed \cite{Banerjee-Bag-Banik-Roy-PRE-2002,Banerjee-Banik-Bag-Ray-PHE-2002,Banik-Bag-Ray-PRE-2002,Banerjee-Bag-Banik-Ray-JCP-2004}. The corresponding
ansatz of a coherent state representation provides a connection with the
classical GLE in the limit of $\hbar\to 0$. However, 
some {\em ad hoc}, not fully justifiable, assumptions 
are used for the form of the quantum Hamiltonian and
the way the quantum thermal averages are performed \cite{Bag-Banik-explanation}.

We also note that in most of the methods mentioned above the bath
Hamiltonian was represented as a set of independent harmonic oscillators. 

Here we offer a fully consistent derivation of cQGLE for an open system
based on the density matrix method. The Born approximation for the
initial density matrix of the whole system is not used, i.e. the whole
system initially is not assumed to be partitioned. We consider a general
Hamiltonian for the open system which is linearly coupled to the harmonic
bath. One of the advantages of our model is that, similarly to our
classical treatment \cite{my-SBC-1,Lorenzo-GLE-2014,Herve-GLE-2015},
the bath and the open system are treated as parts of the same whole
system. We show that, using a plausible so-called ``harmonisation''
approximation, a class of cQGLEs for the mean values $\left\langle x_{i}\right\rangle _{t}$
of atoms in the open quantum system can be established. These equations
are non-Markovian in nature with a friction kernel which is identical
to that found in the classical GLE \cite{my-SBC-1}, while the random
force, contrary to the classical case, is non-Gaussian. It is shown
to have a zero mean with the pair correlation function being of the
same functional form as in the classical case. Next, we establish
a connection with the previously developed methods and obtain an explicit
expression for the pair correlation function of the random force by
assuming that the order in which the stochastic and quantum-mechanical
averages are performed must not affect the final result.

\section{Hamiltonian and exact Heisenberg equations of motion for operators\label{sub:Hamiltonian-and-Heisenber}}

Consider a system consisting of two parts: a finite open system (or
region 1) and an infinite heat bath (region 2). Correspondingly, subscripts
1 and 2 will be used in vectors and matrices, where appropriate. We
assume that the heat bath is much larger than the system itself and
hence can be asigned to have a fixed temperature $T$. The Hamiltonian
of the whole system, $\mathcal{H}=\mathcal{H}_{1}+\mathcal{H}_{2}+\mathcal{H}_{12}$,
contains the Hamiltonian of the open system, 
\begin{equation}
\mathcal{H}_{1}=\sum_{i\in1}\frac{p_{i}^{2}}{2m_{i}}+W\left(\mathbf{x}_{1}\right)=\frac{1}{2}\mathbf{p}_{1}^{T}\mathbf{M}_{11}^{-1}\mathbf{p}_{1}+W\left(\mathbf{x}_{1}\right)\label{eq:H1}
\end{equation}
which assumes an arbitrary potential energy term, $W\left(\mathbf{x}_{1}\right)$,
the harmonic bath,
\begin{equation}
\mathcal{H}_{2}=\frac{1}{2}\sum_{i,j\in2}\left(\frac{q_{i}^{2}}{m_{i}}\delta_{ij}+\Phi_{ij}u_{i}u_{j}\right)=\frac{1}{2}\mathbf{q}_{2}^{T}\mathbf{M}_{22}^{-1}\mathbf{q}_{2}+\frac{1}{2}\mathbf{u}_{2}^{T}\mathbf{\Phi}_{22}\mathbf{u}_{2}\label{eq:H2}
\end{equation}
and the interaction between the system and bath regions which is assumed
to be linear with respect to atomic displcements $u_{k}$ of the bath
atoms: 
\begin{equation}
\mathcal{H}_{12}=\sum_{i\in2}h_{i}u_{i}=\mathbf{h}_{2}^{T}\mathbf{u}_{2}\label{eq:H12}
\end{equation}
Here $\mathbf{x}_{1}=\left(x_{i}\right)$ and $\mathbf{p}_{1}=\left(p_{i}\right)$
are vector-columns of all Cartesian coordinates $i$ of the atoms
in the system and their momenta, respectively; $\mathbf{u}_{2}=\left(u_{i}\right)$
and $\mathbf{q}_{2}=\left(q_{i}\right)$ are vector-columns of all
atomic displacements in the bath and their corresponding momentum
operators. $\mathbf{\Phi}_{22}=\left(\Phi_{ij}\right)$ is the bath
force-constant matrix, and $\mathbf{M}_{11}=\left(\delta_{ij}m_{i}\right)$
and $\mathbf{M}_{22}=\left(\delta_{ij}m_{i}\right)$ are the diagonal
matrices of atomic masses of the system and bath, respectively. The
superscript $T$ means transpose. The vector $\mathbf{h}_{2}=\left(h_{i}\left(\mathbf{x}_{1}\right)\right)$
defines (minus) forces with which atoms in the system act on the atoms
of the bath; this vector is assumed to have an arbitrary dependence
on $\mathbf{x}_{1}$.

The above Hamiltonian is exactly the same as in the recent classical
formulation of the GLE equations \cite{my-SBC-1}. It is more general
than the Caldeira-Leggett Hamiltonian \cite{Caldeira-Leggett-1983}
containing independent harmonic oscillators in the bath and the coupling
which is linear in both bath and system coordinates (the bilinear
coupling), that is normally used in quantum theories of GLE \cite{FKM-J.Math.Ph.-1965,Benguria-Kac-1981,Lindenberg-West-1984,Ford-Kac-JST-1987,Segal-Nitzan-Hanggi-2003,Wang-PRL-2007,Gardiner-book_noise-2010,Weiss-2012}.
Our Hamiltonian can be obtained from the full Hamiltonian of the whole
combined system by expanding it until the second order in terms of
atomic displacements in the bath. Hence, its parameters can be taken
directly from the adopted Hamiltonian of the realistic system under
study \cite{Herve-GLE-2015}. Note that $\mathcal{H}_{1}$ includes
the interaction between atoms in the system and their counterparts
in the bath which are clumped at their equilibrium positions; any
variation of the system-bath interaction due to the bath atoms vibrating
around these positions is accounted for by the rest of the combined
bath Hamiltonian $\mathcal{H}_{b}=\mathcal{H}_{2}+\mathcal{H}_{12}$. 

Next we introduce the density matrix operator, $\rho(t)$, for the
system and bath, which satisfies the appropriate Liouville equation
with the full Hamiltonian. We recall \cite{Zubarev-1} that the general
solution of the Liouville equation for time-independent Hamiltonian
$\mathcal{H}$ is $\rho(t)=U\rho^{0}U^{\dagger}$, where $U\equiv U\left(t,t_{0}\right)$
and $U\left(t_{1},t_{2}\right)=\exp\left[-i\mathcal{H}\left(t_{1}-t_{2}\right)/\hbar\right]$
is the evolution operator, and $\rho^{0}$ is the density matrix at
the initial time $t_{0}$. Correspondingly, an operator $A$ in the
Heisenberg picture (to be denoted with the tilda in the following),
$\widetilde{A}(t)=U^{\dagger}A\,U$, satisfies the equation of motion
$i\hbar\partial_{t}\widetilde{A}(t)=U^{\dagger}\left[A,\mathcal{H}\right]U$. 

Our goal is to obtain a closed set of equations for the expectation
values of the atomic positions, $\left\langle x_{i}\right\rangle _{t}=\mbox{Tr}\left(\rho(t)x_{i}\right)=\mbox{Tr}\left(\rho_{0}\widetilde{x}_{i}(t)\right)$
for $i\in1$, by eliminating the degrees of freedom of the bath atoms.
Ideally, we would like these equations to resemble classical GLE with
a friction memory term and stochastic forces. To this end, instead
of the operators of the bath $\mathbf{u}_{2}$ and $\mathbf{q}_{2}$,
it is convenient to introduce their mass-scaled counterparts $\mathbf{x}_{2}=\left(x_{i};\:i\in2\right)=\mathbf{M}_{22}^{1/2}\mathbf{u}_{2}$
and $\mathbf{p}_{2}=\left(p_{i};\:i\in2\right)=\mathbf{M}_{22}^{-1/2}\mathbf{q}_{2}$,
which satisfy the same commutation relations, $\left[x_{i},p_{j}\right]=i\hbar\delta_{ij}$
($i,j\in2$). Then the combined bath and interaction Hamiltonian takes
on the following form:
\begin{equation}
\mathcal{H}_{b}=\mathcal{H}_{12}+\mathcal{H}_{2}=\frac{1}{2}\mathbf{p}{}_{2}^{T}\mathbf{p}_{2}+\frac{1}{2}\mathbf{x}_{2}^{T}\mathbf{D}_{22}\mathbf{x}_{2}+\mathbf{V}_{2}^{T}\mathbf{x}_{2}\label{eq:H2+H12-1}
\end{equation}
where $\mathbf{V}_{2}=\mathbf{h}_{2}\mathbf{M}_{22}^{-1/2}$ are the
appropriately rescaled coefficients explicitly depending on $\mathbf{x}_{1}$. 

By calculating commutators of the operators of coordinates and momenta
of both regions ($\mathbf{x}_{1}$, $\mathbf{x}_{2}$, $\mathbf{p}_{1}$
and $\mathbf{p}_{2}$) with the Hamiltonian $\mathcal{H}$, the equations
of motion for the operators $\widetilde{\mathbf{x}}_{1}$, $\widetilde{\mathbf{x}}_{2}$,
$\widetilde{\mathbf{p}}_{1}$ and $\widetilde{\mathbf{p}}_{2}$ in
the Heisenberg representation are obtained. For the system we have:
\begin{equation}
\mathbf{M}_{11}\partial_{t}\widetilde{\mathbf{x}}_{1}=\widetilde{\mathbf{p}}_{1}\;,\quad\partial_{t}\widetilde{\mathbf{p}}_{1}=-\mathbf{h}_{1}\left(\widetilde{\mathbf{x}}_{1}\right)-\mathbf{V}_{12}\left(\widetilde{\mathbf{x}}_{1}\right)\widetilde{\mathbf{x}}_{2}\equiv\mathbf{M}_{11}\partial_{t}^{2}\widetilde{\mathbf{x}}_{1}\label{eq:EoM-for-1-1}
\end{equation}
where $\mathbf{h}_{1}\left(\mathbf{x}_{1}\right)=\partial W/\partial\mathbf{x}_{1}=\left(\partial W/\partial x_{i};\;i\in1\right)$
and 
\[
\mathbf{V}_{12}\left(x_{1}\right)=\frac{\partial\mathbf{V}_{2}\left(\mathbf{x}_{1}\right)}{\partial\mathbf{x}_{1}}=\left(\frac{1}{\sqrt{m_{j}}}\frac{\partial h_{j}\left(\mathbf{x}_{1}\right)}{\partial x_{i}};\;i\in1,j\in2\right)
\]
and hence $\mathbf{h}_{1}\left(\widetilde{\mathbf{x}}_{1}\right)=U^{\dagger}\left(\partial W/\partial\mathbf{x}_{1}\right)U=\partial W\left(\widetilde{\mathbf{x}}_{1}\right)/\partial\widetilde{\mathbf{x}}_{1}$
and $\mathbf{V}_{12}\left(\widetilde{\mathbf{x}}_{1}\right)=\partial\mathbf{V}_{2}\left(\widetilde{\mathbf{x}}_{1}\right)/\partial\widetilde{\mathbf{x}}_{1}$. 

Similarly for the bath: 
\begin{equation}
\partial_{t}\widetilde{\mathbf{x}}_{2}=\widetilde{\mathbf{p}}_{2}\;,\quad\partial_{t}\widetilde{\mathbf{p}}_{2}=-\mathbf{D}_{22}\widetilde{\mathbf{x}}_{2}-\mathbf{V}_{2}\left(\widetilde{\mathbf{x}}_{1}\right)\equiv\partial_{t}^{2}\widetilde{\mathbf{x}}_{2}\label{eq:EoM-for-2-1}
\end{equation}
where $\mathbf{D}_{22}=\mathbf{M}_{22}^{-1/2}\mathbf{\Phi}_{22}\mathbf{M}_{22}^{-1/2}$
is the dynamical matrix of the bath. 

The equations (\ref{eq:EoM-for-2-1}) for the coordinates $\widetilde{\mathbf{x}}_{2}(t)$
of the bath atoms are solved in exactly the same way as in the classical
case \cite{my-SBC-1} by first defining normal coordinates $\xi_{\lambda}=\sum_{i\in2}e_{\lambda i}x_{i}=\mathbf{e}_{\lambda}^{T}\mathbf{x}_{1}$
of the bath expressed via the eigenvectors $\mathbf{e}_{\lambda}$
of the dynamical matrix, $\mathbf{D}_{22}\mathbf{e}_{\lambda}=\omega_{\lambda}^{2}\mathbf{e}_{\lambda}$,
where $\omega_{\lambda}$ are frequencies of the bath's normal vibrational
modes. In the new coordinates we obtain decoupled differential equations
for each normal mode $\lambda$ as $\ddot{\widetilde{\xi}}_{\lambda}+\omega_{\lambda}^{2}\widetilde{\xi}_{\lambda}=-V_{\lambda}(t)$
(dots above the symbols denote time derivatives), where $V_{\lambda}(t)=\mathbf{e}_{\lambda}^{T}\mathbf{V}_{2}$,
so that their solutions are readily obtained ($i\in2$):
\[
\widetilde{x}_{i}(t)=\sum_{\lambda}e_{\lambda i}\widetilde{\xi}_{\lambda}(t)=\sum_{\lambda}e_{\lambda i}\left[A_{\lambda}e^{i\omega_{\lambda}t}+B_{\lambda}e^{-i\omega_{\lambda}t}-\frac{1}{\omega_{\lambda}}\int_{t_{0}}^{t}V_{\lambda}\left(\tau\right)\sin\left[\omega_{\lambda}\left(t-\tau\right)\right]d\tau\right]
\]
\[
\widetilde{p}_{i}(t)=\sum_{\lambda}e_{\lambda i}\dot{\widetilde{\xi}}_{\lambda}(t)=\sum_{\lambda}e_{\lambda i}\left[i\omega_{\lambda}\left(A_{\lambda}e^{i\omega_{\lambda}t}-B_{\lambda}e^{-i\omega_{\lambda}t}\right)-\int_{t_{0}}^{t}V_{\lambda}\left(\tau\right)\cos\left[\omega_{\lambda}\left(t-\tau\right)\right]d\tau\right]
\]
where $A_{\lambda}$ and $B_{\lambda}$ are two operators to be determined
from the initial conditions: $\widetilde{x}_{i}(t_{0})=x_{i}$ and
$\widetilde{p}_{i}(t_{0})=p_{i}$. Some simple algebra yields the
following expression for the (rescaled) atomic positions of the bath
atoms: 
\begin{equation}
\mathbf{\widetilde{x}}_{2}(t)=\dot{\mathbf{\Omega}}_{22}\left(t-t_{0}\right)\mathbf{x}_{2}+\mathbf{\Omega}_{22}\left(t-t_{0}\right)\mathbf{p}_{2}-\int_{t_{0}}^{t}\mathbf{\Omega}_{22}\left(t-\tau\right)\mathbf{V}_{2}\left(\tau\right)d\tau\label{eq:x2-tilda-1}
\end{equation}
where 
\begin{equation}
\mathbf{\Omega}_{22}(t)=\sum_{\lambda}\frac{\mathbf{e}_{\lambda}\mathbf{e}_{\lambda}^{T}}{\omega_{\lambda}}\sin\left(\omega_{\lambda}t\right)\label{eq:omega-matrix}
\end{equation}
and 
\begin{equation}
\dot{\mathbf{\Omega}}_{22}(t)=\sum_{\lambda}\mathbf{e}_{\lambda}\mathbf{e}_{\lambda}^{T}\cos\left(\omega_{\lambda}t\right)\label{eq:omega-dot-matrix}
\end{equation}
are two square bath matrices (cf. \cite{my-SBC-1}). The time integral
in Eq. (\ref{eq:x2-tilda-1}) can be calculated by parts. Defining
one more bath matrix \cite{my-SBC-1} 
\begin{equation}
\mathbf{\Pi}_{22}(t)=\sum_{\lambda}\frac{\mathbf{e}_{\lambda}\mathbf{e}_{\lambda}^{T}}{\omega_{\lambda}^{2}}\cos\left(\omega_{\lambda}t\right)\label{eq:pol-matirx}
\end{equation}
and noticing that $\mathbf{D}_{22}^{-1}=\sum_{\lambda}\omega_{\lambda}^{-2}\mathbf{e}_{\lambda}\mathbf{e}_{\lambda}^{T}\equiv\mathbf{\Pi}_{22}(0)$,
we obtain:
\[
\widetilde{\mathbf{x}}_{2}(t)=\left[\dot{\mathbf{\Omega}}_{22}\left(t-t_{0}\right)\mathbf{x}_{2}+\mathbf{\Omega}_{22}\left(t-t_{0}\right)\mathbf{p}_{2}+\mathbf{\Pi}_{22}\left(t-t_{0}\right)\mathbf{V}_{2}\left(t_{0}\right)\right]-\mathbf{D}_{22}^{-1}\mathbf{V}_{2}\left(t\right)
\]
\[
+\int_{t_{0}}^{t}\mathbf{\Pi}_{22}\left(t-\tau\right)\left[\frac{d}{d\tau}\mathbf{V}_{2}\left(\tau\right)\right]d\tau
\]
The time derivative of the operator $\mathbf{V}_{2}\left(\tau\right)$
is 
\begin{equation}
\partial_{\tau}\mathbf{V}_{2}\left(\tau\right)=\frac{1}{i\hbar}U^{\dagger}\left[\mathbf{V}_{2}\left(\mathbf{x}_{1}\right),\mathcal{H}\right]U=\frac{1}{i\hbar}U^{\dagger}\left[\mathbf{V}_{2}\left(\mathbf{x}_{1}\right),\frac{1}{2}\mathbf{p}_{1}^{T}\mathbf{M}_{11}^{-1}\mathbf{p}_{1}\right]U=\mathbf{V}_{21}\left(\mathbf{x}_{1}\right)\partial_{\tau}\widetilde{\mathbf{x}}_{1}\left(\tau\right)-\frac{i\hbar}{2}\bar{\mathbf{V}}_{2}\left(\tau\right)\label{eq:iterm1}
\end{equation}
where $\bar{\mathbf{V}}_{2}\left(\tau\right)=\left(\bar{V}_{j}\left(\tau\right);\,j\in2\right)$
with $\bar{V}_{j}\left(\tau\right)=\sum_{i\in1}m_{i}^{-1}\partial^{2}V_{j}\left(\widetilde{\mathbf{x}}_{1}\right)/\partial\widetilde{x}_{i}^{2}$.
Note that the last term in Eq. (\ref{eq:iterm1}) vanishes in the
$\hbar\rightarrow0$ limit. As we shall see immediately, it will be
responsible for a contribution to the force which does not have the
form of the friction force. Correspondingly, 
\[
\widetilde{\mathbf{x}}_{2}(t)=\left[\dot{\mathbf{\Omega}}_{22}\left(t-t_{0}\right)\mathbf{x}_{2}+\mathbf{\Omega}_{22}\left(t-t_{0}\right)\mathbf{p}_{2}+\mathbf{\Pi}_{22}\left(t-t_{0}\right)\mathbf{V}_{2}\left(t_{0}\right)\right]-\mathbf{D}_{22}^{-1}\mathbf{V}_{2}\left(t\right)
\]
\[
+\int_{t_{0}}^{t}\mathbf{\Pi}_{22}\left(t-\tau\right)\mathbf{V}_{21}\left(\tau\right)\partial_{\tau}\widetilde{\mathbf{x}}_{1}\left(\tau\right)d\tau-\frac{i\hbar}{2}\int_{t_{0}}^{t}\mathbf{\Pi}_{22}\left(t-\tau\right)\bar{\mathbf{V}}_{2}\left(\tau\right)d\tau
\]
Substituting this expression into the equation of motion (\ref{eq:EoM-for-1-1})
for the system atoms, we obtain the quantum GLE (the differential
equation for Heisenberg position operators of the open system):
\[
\mathbf{M}_{11}\ddot{\widetilde{\mathbf{x}}}_{1}(t)=-\mathbf{h}_{1}(t)+\mathbf{V}_{12}(t)\mathbf{D}_{22}^{-1}\mathbf{V}_{2}(t)+\mathbf{R}_{1}(t)
\]
\begin{equation}
-\int_{t_{0}}^{t}\mathbf{K}_{11}\left(t,\tau\right)\dot{\widetilde{\mathbf{x}}}_{1}(\tau)d\tau+\frac{i\hbar}{2}\int_{t_{0}}^{t}\mathbf{V}_{12}\left(t\right)\mathbf{\Pi}_{22}\left(t-\tau\right)\bar{\mathbf{V}}_{2}\left(\tau\right)d\tau\label{eq:general-operator-QGLE}
\end{equation}
where 
\begin{equation}
\mathbf{K}_{11}\left(t,\tau\right)=\mathbf{V}_{12}\left(t\right)\mathbf{\Pi}_{22}\left(t-\tau\right)\mathbf{V}_{21}\left(\tau\right)\label{eq:friction-kernel}
\end{equation}
is the friction kernel (cf. \cite{my-SBC-1}) and 
\begin{equation}
\mathbf{R}_{1}(t)=-\mathbf{V}_{12}(t)\left[\dot{\mathbf{\Omega}}_{22}\left(t-t_{0}\right)\mathbf{x}_{2}+\mathbf{\Omega}_{22}\left(t-t_{0}\right)\mathbf{p}_{2}+\mathbf{\Pi}_{22}\left(t-t_{0}\right)\mathbf{V}_{2}\left(t_{0}\right)\right]\label{eq:random-force1-1}
\end{equation}
The obtained (operator) quantum GLE is exact for the Hamiltonian we
adopted. We first note that the first integral term in the GLE (\ref{eq:general-operator-QGLE})
has exactly the same form as the classical friction force \cite{my-SBC-1};
however, there is also a second integral term, which does not have
the form of a friction force. This is a purely ``non-classical''
term as it disappears in the classical limit of $\hbar\rightarrow0$.
More importantly, the obtained GLE is highly non-linear with respect
to the Heisenberg positions $\widetilde{\mathbf{x}}_{1}\left(t\right)$
as $\mathbf{h}_{1}$, $\mathbf{V}_{2}$, $\mathbf{V}_{21}$ and $\bar{\mathbf{V}}_{2}$
all depend on it (and hence on time). Therefore, when multiplying
both sides of this equation by the initial density matrix $\rho^{0}$
and taking the trace over the Hilbert space of the whole combined
system, it will not be possible to obtain a self-containing equation
for the averages $\left\langle \mathbf{x}_{1}\right\rangle _{t}$.
This is because the average of a function is not generally equal to
the function of the average, e.g. $\left\langle \mathbf{h}_{1}\left(\widetilde{\mathbf{x}}_{1}\right)\right\rangle _{t}=\mbox{Tr}\left[\rho^{0}\mathbf{h}_{1}\left(\widetilde{\mathbf{x}}_{1}\right)\right]\neq\mathbf{h}_{1}\left(\left\langle \mathbf{x}_{1}\right\rangle _{t}\right)$.
Only for \emph{linear} operators $\mathbf{h}_{1}\left(x_{1}\right)=\mathbf{h}_{1}^{0}+\mathbf{h}_{11}^{0}\mathbf{x}_{1}$
we would have the equality, $\left\langle \mathbf{h}_{1}\left(\widetilde{\mathbf{x}}_{1}\right)\right\rangle _{t}=\mathbf{h}_{1}\left(\left\langle \mathbf{x}_{1}\right\rangle _{t}\right)$.
Therefore, there is a certain difficulty in formulating a self-contained
$c-$number quantum GLE in this rather general case. This situation
is fully resolved within the \emph{harmonisation approximation} to
be introduced next.

\begin{figure}
\begin{centering}
\includegraphics[height=2.5cm]{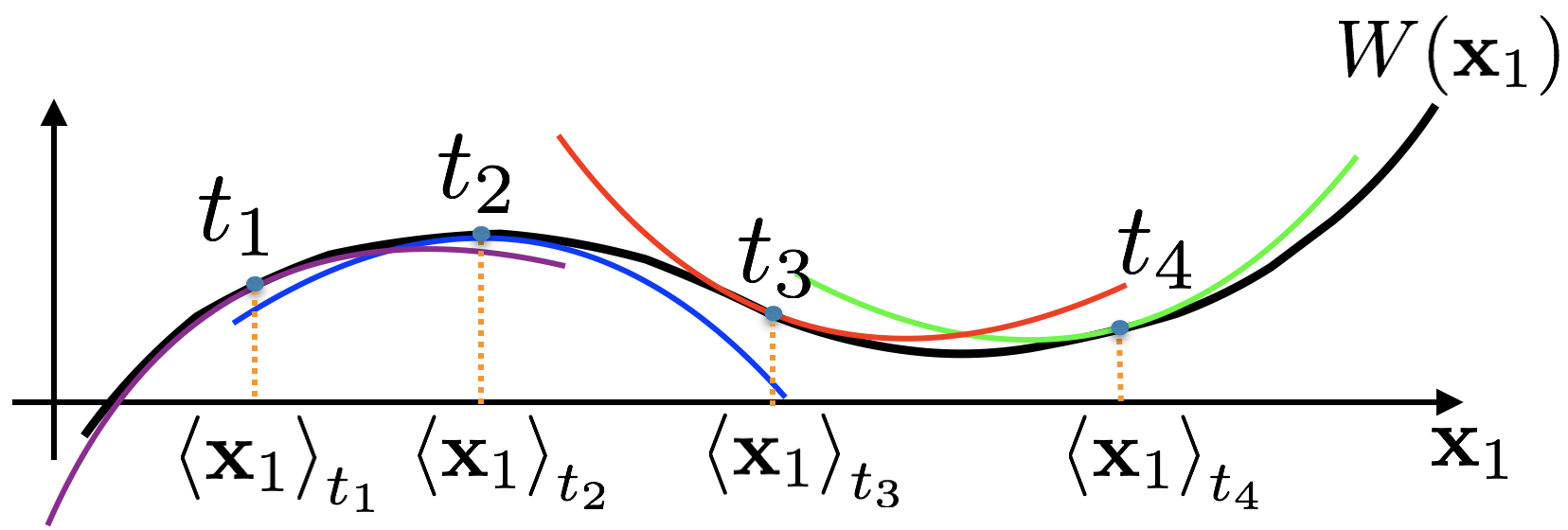} 
\par\end{centering}

\caption{In the harmonisation approximation a general potential function $W\left(\mathbf{x}_{1}\right)$
(black) of the coordinates $\mathbf{x}_{1}$ is approximated by parabolas
at different times (coloured curves) which are fit around the exact
averages $\left\langle \mathbf{x}_{1}\right\rangle _{t}$. \label{fig:harmonisation}}
\end{figure}

\section{Quantum GLE within the harmonisation approximation}

In order to obtain a closed set of equations for the expectation values
of the atomic positions, $\left\langle x_{i}\right\rangle _{t}$ for
all $i\in1$, we make what we shall call a \emph{harmonisation approximation}
(cf. \cite{CEID-JPCM-2005}), whereby $W\left(\mathbf{x}_{1}\right)$
and $\mathbf{h}_{2}\left(\mathbf{x}_{1}\right)$ terms in $\mathcal{H}_{1}$
and $\mathcal{H}_{12}$, respectively, are expanded in terms of the
displacements $u_{i}=x_{i}-\left\langle x_{i}\right\rangle _{t}$
of atoms in the open system with respect to their \emph{instantaneous}
positions $\left\langle x_{i}\right\rangle _{t}$ up to quadratic
terms:

\begin{equation}
\mathcal{H}_{1}\simeq\frac{1}{2}\mathbf{p}_{1}^{T}\mathbf{M}_{11}^{-1}\mathbf{p}_{1}+W\left(t\right)+\mathbf{h}_{1}^{T}(t)\mathbf{u}_{1}+\frac{1}{2}\mathbf{u}_{1}^{T}\mathbf{\Phi}_{11}(t)\mathbf{u}_{1}\label{eq:H1-expnaded}
\end{equation}
\begin{equation}
\mathcal{H}_{12}=\mathbf{h}_{2}^{T}(t)\mathbf{u}_{2}+\mathbf{u}_{1}^{T}\mathbf{h}_{12}(t)\mathbf{u}_{2}\label{eq:H12-expanded}
\end{equation}
where expansion coefficients are: $W=W\left(\left\langle \mathbf{x}_{1}\right\rangle _{t}\right)$,
$\mathbf{h}_{2}(t)=\left(h_{i}\left(\left\langle \mathbf{x}_{1}\right\rangle _{t}\right);\;i\in2\right)$,
\[
\mathbf{h}_{1}(t)=\left(h_{i};\;i\in1\right)=\frac{\partial W\left(\left\langle \mathbf{x}_{1}\right\rangle _{t}\right)}{\partial\left\langle \mathbf{x}_{1}\right\rangle _{t}}\quad\mbox{with}\quad h_{i}=\frac{\partial W}{\partial\left\langle x_{i}\right\rangle _{t}}\;,
\]
\[
\mathbf{\Phi}_{11}(t)=\left(\Phi_{ij};\;i,j\in1\right)=\frac{\partial^{2}W\left(\left\langle \mathbf{x}_{1}\right\rangle _{t}\right)}{\partial\left\langle \mathbf{x}_{1}\right\rangle _{t}\partial\left\langle \mathbf{x}_{1}\right\rangle _{t}}\quad\mbox{with}\quad\Phi_{ij}=\frac{\partial W}{\partial\left\langle x_{i}\right\rangle _{t}\partial\left\langle x_{j}\right\rangle _{t}}\;,
\]
and 
\[
\mathbf{h}_{12}(t)=\left(h_{ij};\;i\in1,j\in2\right)=\frac{\partial\mathbf{h}_{2}\left(\left\langle \mathbf{x}_{1}\right\rangle _{t}\right)}{\partial\left\langle \mathbf{x}_{1}\right\rangle _{t}}\quad\mbox{with}\quad h_{ij}=\frac{\partial h_{j}\left(\left\langle \mathbf{x}_{1}\right\rangle _{t}\right)}{\partial\left\langle x_{i}\right\rangle _{t}}
\]
All of the above expressions are ordinary derivatives of the interaction
$W$ and the (minus) forces $h_{2}$, both being $c-$numbers, i.e.
they are to be understood as real functions of the exact\emph{ instantaneous}
atomic positions $\left\langle \mathbf{x}_{1}\right\rangle _{t}$.
Note that all these coefficients depend on these averages and hence
become explicit functions of time; they are not operators and can
be easily calculated given the explicit functional dependences in
$W\left(\mathbf{x}_{1}\right)$ and $\mathbf{h}_{2}\left(\mathbf{x}_{1}\right)$. 

It is important to realise that the harmonisation approximation goes
beyond the usual harmonic approximation in which the interactions
in the Hamiltonian are expanded with respect to \emph{fixed }atomic
positions. The difference is illustarted in Fig. \ref{fig:harmonisation}
for the interaction $W\left(\mathbf{x}_{1}\right)$. As the time evolves,
the expectation value of the atomic positions changes and the expansion
of the Hamiltonian changes with them. Of course, at this stage we
do not know what the atomic positions $\left\langle \mathbf{x}_{1}\right\rangle _{t}$
are; our goal is to establish a closed equation of motion which would
enable us to determine them.

First, an equation of motion for the operators needs to be obtained.
The method of the previous section can be applied here with the caviat
that now, after the harmonisation approximation, the Hamiltonian depends
on time explicitly, $\mathcal{H\equiv\mathcal{H}}(t)$, since $\mathcal{H}_{1}$
and $\mathcal{H}_{12}$ from Eqs. (\ref{eq:H1-expnaded}) and (\ref{eq:H12-expanded}),
respectively, do. Hence the time evolution of the density matrix and
the Heisenberg representation of operators are to be obtained using
a more general evolution operator 
\[
U\left(t,t_{0}\right)=\widehat{T}\exp\left(-\frac{i}{\hbar}\int_{t_{0}}^{t}\mathcal{H}\left(\tau\right)d\tau\right)
\]
where $\widehat{T}$ is the time-ordering operator (assuming $t>t_{0}$).
The method of obtaining the equations of motion for the operators
is still straightforward as it requires the calculation of the commutators
of the operators $\mathbf{x}_{1}$, $\mathbf{p}_{1}$, $\mathbf{q}_{2}$
and $\mathbf{u}_{2}$ with the Hamiltonian. 

Introducing again the rescaled variables for the bath atoms, the combined
bath and interaction Hamiltonian takes on a simpler form: 
\begin{equation}
\mathcal{H}_{b}=\mathcal{H}_{12}+\mathcal{H}_{2}=\frac{1}{2}\mathbf{p}{}_{2}^{T}\mathbf{p}_{2}+\frac{1}{2}\mathbf{x}_{2}^{T}\mathbf{D}_{22}\mathbf{x}_{2}+\left(\mathbf{V}_{2}+\mathbf{V}_{21}\mathbf{u}_{1}\right)^{T}\mathbf{x}_{2}\label{eq:H2+H12}
\end{equation}
where $\mathbf{V}_{2}=\mathbf{h}_{2}\mathbf{M}_{22}^{-1/2}$ and $\mathbf{V}_{12}=\mathbf{V}_{21}^{T}=\mathbf{h}_{12}\mathbf{M}_{22}^{-1/2}$
are the appropriately rescaled coefficients (which depend on the averages
$\left\langle \mathbf{x}_{1}\right\rangle _{t}$). Calculating the
commutators with $\mathcal{H}$, the following equations of motion
are obtained for the operators in the Heisenberg picture:
\begin{equation}
\mathbf{M}_{11}\partial_{t}\widetilde{\mathbf{x}}_{1}=\widetilde{\mathbf{p}}_{1}\;,\quad\partial_{t}\widetilde{\mathbf{p}}_{1}=-\mathbf{h}_{1}-\mathbf{V}_{12}\widetilde{\mathbf{x}}_{2}-\mathbf{\Phi}_{11}\widetilde{\mathbf{u}}_{1}\equiv\mathbf{M}_{11}\partial_{t}^{2}\widetilde{\mathbf{x}}_{1}\label{eq:EoM-for-1}
\end{equation}
and 
\begin{equation}
\partial_{t}\widetilde{\mathbf{x}}_{2}=\widetilde{\mathbf{p}}_{2}\;,\quad\partial_{t}\widetilde{\mathbf{p}}_{2}=-\mathbf{D}_{22}\widetilde{\mathbf{x}}_{2}-\left(\mathbf{V}_{2}+\mathbf{V}_{21}\widetilde{\mathbf{u}}_{1}\right)\equiv\partial_{t}^{2}\widetilde{\mathbf{x}}_{2}\label{eq:EoM-for-2}
\end{equation}
This results in the decoupled differential equations for each normal
mode $\lambda$ as $\ddot{\widetilde{\xi}}_{\lambda}+\omega_{\lambda}^{2}\widetilde{\xi}_{\lambda}=-V_{\lambda}(t)$,
where this time $V_{\lambda}(t)=\mathbf{e}_{\lambda}^{T}\left(\mathbf{V}_{2}+\mathbf{V}_{21}\widetilde{\mathbf{u}}_{1}\right)$.
Note that the non-linear term we encountered in Eq. (\ref{eq:iterm1})
in the previous Section does not appear here, i.e. it is absent within
the harmonisation approximation. 

Then, the equations for the bath are easily solved similarly to the
general case considered in the previous section: 
\begin{equation}
\widetilde{\mathbf{x}}_{2}(t)=\dot{\mathbf{\Omega}}_{22}\left(t-t_{0}\right)\mathbf{x}_{2}+\mathbf{\Omega}_{22}\left(t-t_{0}\right)\mathbf{p}_{2}-\int_{t_{0}}^{t}\mathbf{\Omega}_{22}\left(t-\tau\right)\left[\mathbf{V}_{2}\left(\tau\right)+\mathbf{V}_{21}\left(\tau\right)\widetilde{\mathbf{u}}_{1}(\tau)\right]d\tau\label{eq:x2-tilda-1-1}
\end{equation}
Noticing that $\partial_{\tau}\mathbf{V}_{2}(\tau)=\partial_{\tau}\mathbf{V}_{2}\left(\left\langle \mathbf{x}_{1}\right\rangle _{\tau}\right)=\mathbf{V}_{21}\partial_{\tau}\left\langle \mathbf{x}_{1}\right\rangle _{\tau}$,
the integral above is calculated by parts to yield:
\[
\widetilde{\mathbf{x}}_{2}(t)=\left[\dot{\mathbf{\Omega}}_{22}\left(t-t_{0}\right)\mathbf{x}_{2}+\mathbf{\Omega}_{22}\left(t-t_{0}\right)\mathbf{p}_{2}+\mathbf{\Pi}_{22}\left(t-t_{0}\right)\mathbf{g}_{2}\right]
\]
\begin{equation}
-\mathbf{D}_{22}^{-1}\left[\mathbf{V}_{2}(t)+\mathbf{V}_{21}(t)\widetilde{\mathbf{u}}_{1}(t)\right]+\int_{t_{0}}^{t}\mathbf{\Pi}_{22}\left(t-\tau\right)\left[\dot{\mathbf{V}}_{21}\left(\tau\right)\widetilde{\mathbf{u}}_{1}(\tau)+\mathbf{V}_{21}\left(\tau\right)\dot{\widetilde{\mathbf{x}}}_{1}(\tau)\right]d\tau\label{eq:x2-final}
\end{equation}
where $\mathbf{u}_{1}=\mathbf{x}_{1}-\left\langle \mathbf{x}_{1}\right\rangle _{t_{0}}$
and 
\begin{equation}
\mathbf{g}_{2}=\mathbf{V}_{2}\left(t_{0}\right)+\mathbf{V}_{21}\left(t_{0}\right)\mathbf{u}_{1}\label{eq:g2}
\end{equation}
Substituting this solution into Eq. (\ref{eq:EoM-for-1}), we arrive
at the following differential equation for the position operators
of the system atoms: 
\begin{equation}
\mathbf{M}_{11}\ddot{\widetilde{\mathbf{x}}}_{1}(t)=\mathbf{F}_{1}(t)-\mathbf{L}_{11}(t)\widetilde{\mathbf{u}}_{1}(t)+\mathbf{R}_{1}(t)-\int_{t_{0}}^{t}\mathbf{V}_{12}\left(t\right)\mathbf{\Pi}_{22}\left(t-\tau\right)\left[\dot{\mathbf{V}}_{21}\left(\tau\right)\widetilde{\mathbf{u}}_{1}(\tau)+\mathbf{V}_{21}\left(\tau\right)\dot{\widetilde{\mathbf{x}}}_{1}(\tau)\right]d\tau\label{eq:for-X1}
\end{equation}
where
\begin{equation}
\mathbf{F}_{1}(t)=-\mathbf{h}_{1}(t)+\mathbf{V}_{12}(t)\mathbf{D}_{22}^{-1}\mathbf{V}_{2}(t)\label{eq:gen-force-in1}
\end{equation}
\begin{equation}
\mathbf{L}_{11}(t)=\mathbf{\Phi}_{11}(t)-\mathbf{V}_{12}(t)\mathbf{D}_{22}^{-1}\mathbf{V}_{21}(t)\label{eq:gen-force-const-matrix-in1}
\end{equation}
are the generalised force and the force-constant matrix for atoms
in the open system, respectively. The second terms in the right hand
sides of expressions (\ref{eq:gen-force-in1}) and (\ref{eq:gen-force-const-matrix-in1})
are related to the contribution of the vibrating baths atoms. Finally,
\begin{equation}
\mathbf{R}_{1}(t)=-\mathbf{V}_{12}(t)\left[\dot{\mathbf{\Omega}}_{22}\left(t-t_{0}\right)\mathbf{x}_{2}+\mathbf{\Omega}_{22}\left(t-t_{0}\right)\mathbf{p}_{2}+\mathbf{\Pi}_{22}\left(t-t_{0}\right)\mathbf{g}_{2}\right]\label{eq:random-force1}
\end{equation}
is an operator acting in the Hilbert space of the open system (due
to $\mathbf{g}_{2}$, Eq. (\ref{eq:g2})) and of the bath (due to
$\mathbf{x}_{2}$ and $\mathbf{p}_{2}$). The meaning of this operator
will be clarified later on. All other terms in Eq. (\ref{eq:for-X1})
depend explicitly on the mean values $\left\langle \mathbf{x}_{1}\right\rangle _{t}$
of the atomic positions in the open system and hence are \emph{not}
operators. 

Note that formally equations (\ref{eq:for-X1})-(\ref{eq:random-force1})
correspond to the initial Hamiltonian (\ref{eq:H1})-(\ref{eq:H12}):
even though the harmonisation approximation was used, the parameters
of the Hamiltonian, Eqs. (\ref{eq:H1-expnaded}) and (\ref{eq:H2+H12}),
depend on time according to the actual system dynamics and the shape
of the potential energy terms of the original Hamiltonian.

The obtained equation of motion (\ref{eq:for-X1}) represents what
is sometimes called the quantum GLE and which has been known (for
somewhat simpler Hamiltonians) since the pioneering work of FKM \cite{FKM-J.Math.Ph.-1965,Gardiner-book_noise-2010}.
The main problem associated with this equation is that it is written
for \emph{operators }acting in the whole Hilbert space of the open
system and bath. Hence, determination of the expectation values of
the atomic positions, the quantities which represent the actual interest,
is an additional and rather complex problem. Instead, our objective
here is to derive an equation \emph{directly} for these expectation
values, the so-called $c-$number GLE for the coordinates of the atoms
of the open system, which would enable one to compute average atomic
trajectories $\left\langle \mathbf{x}_{1}\right\rangle _{t}$ as a
function of time taking full account of the bath. To achieve this
goal, we have to multiply both sides of Eq. (\ref{eq:for-X1}) by
the initial density matrix $\rho^{0}$ and then take the trace over
the whole Hilbert space (system+bath). To accomplish this, we need
an explcit expression for the initial density matrix first.

\section{$c-$number quantum GLE\label{sec:Initial-density-matrix}}

We shall start by finding eigenvectors of the bath Hamiltonian (\ref{eq:H2+H12})
at the initial time $t_{0}$ in which the displacements of atoms of
the system, $\widetilde{\mathbf{u}}_{1}(t_{0})=\mathbf{u}_{1}$, are
considered \emph{as parameters}. The Hamiltonian
\begin{equation}
\mathcal{H}_{b}^{0}=\left(\mathcal{H}_{2}+\mathcal{H}_{12}\right)_{t_{0}}=\frac{1}{2}\mathbf{p}{}_{2}^{T}\mathbf{p}_{2}+\frac{1}{2}\mathbf{x}_{2}^{T}\mathbf{D}_{22}\mathbf{x}_{2}+\mathbf{g}_{2}^{T}\mathbf{x}_{2}\label{eq:Bath-H-at-t0}
\end{equation}
corresponds to a set of displaced harmonic oscillators (the bath in
the presence of the open system) and can be diagonalised exactly using
the canonical transformation: 
\[
\mathcal{U}^{\dagger}\mathcal{H}_{b}^{0}\mathcal{U}=\sum_{\lambda}\hbar\omega_{\lambda}\left(b_{\lambda}^{\dagger}b_{\lambda}+\frac{1}{2}\right)+E_{pol}
\]
with $\mathcal{U}=\prod_{\lambda}\mathcal{U}_{\lambda}=\prod_{\lambda}\exp\left(\gamma_{\lambda}b_{\lambda}^{\dagger}-\gamma_{\lambda}b_{\lambda}\right)$
and $E_{pol}=-\sum_{\lambda}g_{\lambda}^{2}/2\omega_{\lambda}^{2}$,
where the constants $\gamma_{\lambda}=-g_{\lambda}/\sqrt{2\hbar}\omega_{\lambda}^{3/2}$
with $g_{\lambda}=\mathbf{e}_{\lambda}^{T}\mathbf{g}_{2}$ were introduced.
Here $b_{\lambda}^{\dagger}$ and $b_{\lambda}$ are phonon creation
and annihilation operators for the mode $\lambda$ satisfying usual
commutation relations for bosons. Therefore, the eigenvectors and
eigenvalues of $\mathcal{H}_{b}^{0}$ are, respectively: $\left|\psi_{n}\right\rangle =\mathcal{U}\left|n\right\rangle =\prod_{\lambda}\mathcal{U}_{\lambda}\left|n_{\lambda}\right\rangle $
and $E_{n}=E_{n}^{0}+E_{pol}$, where $E_{n}^{0}=\sum_{\lambda}\hbar\omega_{\lambda}\left(n_{\lambda}+\frac{1}{2}\right)$,
and 
\[
\left|n\right\rangle =\prod_{\lambda}\left|n_{\lambda}\right\rangle =\prod_{\lambda}\frac{\left(b_{\lambda}^{\dagger}\right)^{n_{\lambda}}}{\sqrt{n_{\lambda}!}}\left|0\right\rangle 
\]
is a product of the eigenstates $\left|n_{\lambda}\right\rangle $
of the undisplaced $\lambda$-oscillators, $n=\left\{ n_{\lambda}\right\} $
is a set of integer numbers $n_{\lambda}=0,1,2,\ldots$ characterising
excitations of each of the oscillators. The states $\left|n\right\rangle $
are orthonormal, $\left\langle n\right.\left|m\right\rangle =\delta_{nm}$.
It is essential to realise that the eigenstates $\left|\psi_{n}\right\rangle $
depend parametrically on the displacements $\mathbf{u}_{1}$ of atoms
in the open system (via $\mathbf{g}_{2}$, Eq. (\ref{eq:g2})).

The initial density matrix $\rho^{0}$ of the whole system, corresponding
to the bath being at equilibrium with the inverse temperature $\beta=1/k_{B}T$,
can be generally written via eigenstates of the Hamiltonian $\mathcal{H}_{b}^{0}$
as follows:
\begin{equation}
\rho^{0}=\sum_{n}\rho_{n}^{eq}\left|\psi_{n}\right\rangle \rho_{1}^{0}\left\langle \psi_{n}\right|\label{eq:Ro_0}
\end{equation}
where $\rho_{n}^{eq}=\frac{1}{Z_{2}^{0}}e^{-\beta E_{n}^{0}}$ is
an eigenvalue of the equilibrium density matrix $\rho_{2}^{eq}=\frac{1}{Z_{2}^{0}}e^{-\beta\mathcal{H}_{2}}$
of (an isolated) bath and $Z_{2}^{0}=\sum_{n}e^{-\beta E_{n}^{0}}$
the corresponding partition function. 

Tracing out the states of the bath should reduce this density matrix
to the density matrix $\rho_{1}^{0}$ of the open system at the initial
time, and with the choice made above, this is indeed the case:
\[
\mbox{Tr}_{2}\left(\rho^{0}\right)=\sum_{m}\left\langle \psi_{m}\right|\rho^{0}\left|\psi_{m}\right\rangle =\sum_{m}\left\langle \psi_{m}\right|\left\{ \sum_{n}\left|\psi_{n}\right\rangle \rho_{n}^{eq}\rho_{1}^{0}\left\langle \psi_{n}\right|\right\} \left|\psi_{m}\right\rangle =\left(\sum_{m}\rho_{m}^{eq}\right)\rho_{1}^{0}=\rho_{1}^{0}
\]
When calculating the trace over the bath, we used the eigenstates
$\left|\psi_{n}\right\rangle $ of $\mathcal{H}_{b}^{0}$. Note also
that $\sum_{m}\rho_{m}^{eq}=1$ due to normalisation of the bath density
matrix $\rho_{2}^{eq}$. 

Note that this expression is partition-free, i.e. it is not based
on the usually invoked Born approximation \cite{Feynman-Vernon-1963,Petuccione-Open-Systems-2007}.
This is because the initial density matrix $\rho_{1}^{0}$ of the
open system depends on the atomic positions there and hence cannot
be taken outside the sum over states $\left|\psi_{n}\right\rangle $
which also depend explicitly on these positions (via ${\bf g}_2$).

In order to introduce a stochastic field into our formulation, we
realise that generally any density matrix can always be expanded in
terms of the eigenstates of the bath: $\rho=\sum_{nm}\left|\psi_{n}\right\rangle \rho_{nm}^{\prime}\left\langle \psi_{m}\right|$,
where $\rho_{nm}^{\prime}$ are operators acting within the Hilbert
space of the open system. This expression is exact and also partition-free.
Inspired by the work of Ref. \cite{Gaspard-JCP-1999}, we choose the
operator-coefficients $\rho_{nm}^{\prime}$ in the initial density
matrix as 
\begin{equation}
\rho_{nm}^{\prime}=\rho_{nm}e^{i\left(\theta_{n}-\theta_{m}\right)}=\rho_{nm}\prod_{\lambda}e^{i\left(\theta_{n_{\lambda}}-\theta_{m_{\lambda}}\right)}\label{eq:ro-0-with-phases}
\end{equation}
where $\theta_{n_{\lambda}}$ are random numbers (phases) \emph{uniformly
distributed} between $0$ and $2\pi$ and $\rho_{nm}$ are operators
acting in the Hilbert space of the open system only. Hence the following
ansatz is proposed for the initial density matrix of the whole system:
\begin{equation}
\rho^{S}=\sum_{nm}e^{i\left(\theta_{n}-\theta_{m}\right)}\left|\psi_{n}\right\rangle \rho_{nm}\left\langle \psi_{m}\right|\label{eq:stochastic-rho-full}
\end{equation}
The density matrix written in this way can be thought of as being
expanded in terms of the bath exact oscillatory functions, $\left|\psi_{n}\right\rangle e^{i\theta_{n}}$,
which contain random phases $\theta_{n}$. We have indicated explicitly
with the superscript $S$ that this density matrix is stochastic in
nature. 

Averages with respect to the random field of phases will be denoted
with the over-bar. Importantly, 
\[
\overline{e^{i\theta_{n}}}=\prod_{\lambda}\overline{e^{i\theta_{n_{\lambda}}}}=\prod_{\lambda}\frac{1}{2\pi}\int_{0}^{2\pi}e^{i\theta_{n_{\lambda}}}d\theta_{n_{\lambda}}=0
\]
so that, when $n\neq m$ (i.e. $n_{\lambda}\neq m_{\lambda}$ for
at least one mode $\lambda$)
\[
\overline{e^{i\left(\theta_{n}-\theta_{m}\right)}}=\left(\prod_{\lambda\in\left\{ n_{\lambda}\neq m_{\lambda}\right\} }\overline{e^{i\theta_{n_{\lambda}}}e^{-i\theta_{m_{\lambda}}}}\right)\left(\prod_{\lambda\in\left\{ n_{\lambda}=m_{\lambda}\right\} }\overline{1}\right)=\prod_{\lambda\in\left\{ n_{\lambda}\neq m_{\lambda}\right\} }\overline{e^{i\theta_{n_{\lambda}}}}\:\overline{e^{i\theta_{m_{\lambda}}}}^{*}=0
\]
while if $n=m$ (i.e. $n_{\lambda}=m_{\lambda}$ for any $\lambda$)
we have $\left[\overline{e^{i\left(\theta_{n}-\theta_{m}\right)}}\right]_{n=m}=\overline{1}=\prod_{\lambda}\frac{1}{2\pi}\int_{0}^{2\pi}d\theta_{n_{\lambda}}=1$.
Hence, generally 
\begin{equation}
\overline{e^{i\left(\theta_{n}-\theta_{m}\right)}}=\prod_{\lambda}\delta_{n_{\lambda}m_{\lambda}}=\delta_{nm}\label{eq:average-of-exp-with-phases}
\end{equation}

To introduce the \emph{temperature} into our description, we postulate
that the \emph{stochastic field average} of $\rho^{S}$, 
\begin{equation}
\overline{\rho^{S}}=\sum_{nm}\overline{e^{i\left(\theta_{n}-\theta_{m}\right)}}\left|\psi_{n}\right\rangle \rho_{nm}\left\langle \psi_{m}\right|=\sum_{n}\left|\psi_{n}\right\rangle \rho_{nn}\left\langle \psi_{n}\right|\label{eq:equil-rho-full}
\end{equation}
coincides with the exact initial density matrix of Eq. (\ref{eq:Ro_0}):
$\overline{\rho^{S}}\equiv\rho^{0}$. This procedure sets up only
diagonal elements of the operators as $\rho_{nn}=\rho_{1}^{0}\rho_{n}^{eq}$;
non-diagonal operators $\rho_{nm}$ still remain undetermined at this
stage. Note that $\mbox{Tr}_{2}\left(\rho^{S}\right)=\sum_{n}\rho_{nn}=\rho_{1}^{0}$. 

To obtain an equation for the exact averages $\left\langle \mathbf{x}_{1}\right\rangle _{t}^{S}$
for atoms in the open system, we multiply both sides of Eq. (\ref{eq:for-X1})
by $\rho^{S}$ from (\ref{eq:stochastic-rho-full}) and then take
the trace over the whole Hilbert space (using eigenstates of $\mathcal{H}_{b}^{0}$).
The superscript $S$ in $\left\langle \mathbf{x}_{1}\right\rangle _{t}^{S}=\mbox{Tr}\left(\rho^{S}\widetilde{\mathbf{x}}_{1}(t)\right)$
indicates that a particular manifestation of the stochastic field
(a particular set of random phases) is used. Note that, when calculating
a given trajectory, the harmonisation approximation is made with respect
to these particular averages corresponding to the given realisation
of the stochastic field. Then, multiplying both sides of Eq. (\ref{eq:for-X1})
by $\rho^{S}$, taking the trace and noting that $\mbox{Tr}\left(\rho^{S}\widetilde{\mathbf{u}}_{1}(t)\right)=\mbox{Tr}\left(\rho^{S}\widetilde{\mathbf{x}}_{1}(t)\right)-\left\langle \mathbf{x}_{1}\right\rangle _{t}^{S}=0$
for any time, we obtain the desired cQGLE: 
\begin{equation}
\mathbf{M}_{11}\partial_{t}^{2}\left\langle \mathbf{x}_{1}\right\rangle _{t}^{S}=\mathcal{\mathbf{F}}_{1}(t)+\mathbf{\mathbfcal{R}}_{1}(t)-\int_{t_{0}}^{t}\mathbf{K}_{11}\left(t,\tau\right)\partial_{t}\left\langle \mathbf{x}_{1}\right\rangle _{\tau}^{S}d\tau\label{eq:QGLE-1}
\end{equation}
where $\mathbf{K}_{11}(t,\tau)$ is the friction kernel (\ref{eq:friction-kernel}),
$\mathcal{\mathbf{F}}_{1}(t)$ is the conservative force (\ref{eq:gen-force-in1})
containing a ``polaron''-like contribution from the bath (cf. \cite{my-SBC-1,Lorenzo-GLE-2014})
elastically responding to the atomic positions in the open system
and 
\begin{equation}
\mathbfcal{R}_{1}(t)=\mbox{Tr}\left(\rho^{S}\mathbf{R}_{1}(t)\right)=\mbox{Tr}_{1}\left\langle \mathbf{R}_{1}(t)\right\rangle _{2}^{S}\label{eq:av-of-R1-ini}
\end{equation}
is the random force, where $\left\langle \ldots\right\rangle _{2}^{S}=\mbox{Tr}_{2}\left(\rho^{S}\ldots\right)$
is the statistical average over the bath, while $\mbox{Tr}_{1}$ corresponds
to the trace over the states of the open system. 

Importantly, the derived cQGLE is self-contained as all of the time-dependent
terms in it are explicit functions of the averages $\left\langle \mathbf{x}_{1}\right\rangle _{t}^{S}$
which this equation defines. At first site, the cQGLE has the same
form as the classical GLE \cite{my-SBC-1,Lorenzo-GLE-2014}. However,
the behaviour of the random force, as will be shown below, is very
different from the classical case.

\section{Properties of the random force}

The random force (\ref{eq:av-of-R1-ini}) contains the random phases
associated with all harmonic oscillators of the bath; at the same
time, it depends explcitily on time and hence represents a \emph{stochastic
process}. To define explicitly the cQGLE, it is necessary to study
this stochastic process in more detail. In particular, we would like
to establish whether the random force $\mathbfcal{R}_{1}(t)=\left(\mathcal{R}_{i}(t);\:i\in1\right)$
is a Gaussian or a non-Gaussian process. It is known \cite{Risken}
that if the process is Gaussian, then odd-moment correlation functions
must be equal to zero, while even-moment correlation functions must
be equal to a sum of products of all pair correlation functions. We
shall explicitly show here that if the first statement appears to
be true, the second one is not, proving that the stochastic field,
which we have introduced above, in not Gaussian.

\subsection{Random force}

To calculate the random force, we first take the trace of the operator
(\ref{eq:random-force1-1}) over the bath states: 

\begin{equation}
\left\langle \mathbf{R}_{1}(t)\right\rangle _{2}^{S}=-\mathbf{V}_{12}(t)\left[\dot{\mathbf{\Omega}}_{22}\left(t-t_{0}\right)\left\langle \mathbf{x}_{2}\right\rangle _{2}^{S}+\mathbf{\Omega}_{22}\left(t-t_{0}\right)\left\langle \mathbf{p}_{2}\right\rangle _{2}^{S}+\mathbf{\Pi}_{22}\left(t-t_{0}\right)\left\langle \mathbf{g}_{2}\right\rangle _{2}^{S}\right]\label{eq:bath-avergage-R1}
\end{equation}
where 
\begin{equation}
\left\langle \mathbf{x}_{2}\right\rangle _{2}^{S}=\sum_{mn}\left\langle \psi_{m}\right|\mathbf{x}_{2}\left|\psi_{n}\right\rangle \rho_{nm}e^{i\left(\theta_{n}-\theta_{m}\right)}\label{eq:X2-interm1}
\end{equation}
\begin{equation}
\left\langle \mathbf{p}_{2}\right\rangle _{2}^{S}=\sum_{mn}\left\langle \psi_{m}\right|\mathbf{p}_{2}\left|\psi_{n}\right\rangle \rho_{nm}e^{i\left(\theta_{n}-\theta_{m}\right)}\label{eq:P2-interm1-1}
\end{equation}
\begin{equation}
\left\langle \mathbf{g}_{2}\right\rangle _{2}^{S}=\left\langle \mathbf{V}_{2}\left(t_{0}\right)+\mathbf{V}_{21}\left(t_{0}\right)\mathbf{u}_{1}\right\rangle _{2}^{S}=\rho_{1}^{0}\left(\mathbf{V}_{2}\left(t_{0}\right)+\mathbf{V}_{21}\left(t_{0}\right)\mathbf{u}_{1}\right)=\rho_{1}^{0}\mathbf{g}_{2}\label{eq:G2}
\end{equation}
The matrix elements $\left\langle \psi_{m}\right|\mathbf{x}_{2}\left|\psi_{n}\right\rangle $
and $\left\langle \psi_{m}\right|\mathbf{p}_{2}\left|\psi_{n}\right\rangle $
which are needed for calculating the averages $\left\langle \mathbf{x}_{2}\right\rangle _{2}$
and $\left\langle \mathbf{p}_{2}\right\rangle _{2}$, are obtained
by making use of the explicit expressions for the position and momenta
operators (written via bath modes creation and annihilation operators),
\begin{equation}
\mathbf{x}_{2}=\sum_{\lambda}\sqrt{\frac{\hbar}{2\omega_{\lambda}}}\mathbf{e}_{\lambda}\left(b_{\lambda}^{\dagger}+b_{\lambda}\right)\;,\label{eq:x2-via-bosons}
\end{equation}
\begin{equation}
\mathbf{p}_{2}=i\sum_{\lambda}\sqrt{\frac{\hbar\omega_{\lambda}}{2}}\mathbf{e}_{\lambda}\left(b_{\lambda}^{\dagger}-b_{\lambda}\right)\;,\label{eq:p2-via-bosons}
\end{equation}
the fact that $\left|\psi_{n}\right\rangle =\mathcal{U}\left|n\right\rangle $,
and also that $\mathcal{U}^{\dagger}b_{\lambda}\mathcal{U}=b_{\lambda}+\gamma_{\lambda}$
and $\mathcal{U}^{\dagger}b_{\lambda}^{\dagger}\mathcal{U}=b_{\lambda}^{\dagger}+\gamma_{\lambda}$.
We have: 
\[
\left\langle \psi_{m}\right|\mathbf{x}_{2}\left|\psi_{n}\right\rangle =\sum_{\lambda}\sqrt{\frac{\hbar}{2\omega_{\lambda}}}\mathbf{e}_{\lambda}\left\langle m\right|\mathcal{U}^{\dagger}\left(b_{\lambda}^{\dagger}+b_{\lambda}\right)\mathcal{U}\left|n\right\rangle 
\]
\[
=\delta_{nm}\sum_{\lambda}\sqrt{\frac{2\hbar}{\omega_{\lambda}}}\mathbf{e}_{\lambda}\gamma_{\lambda}+\sum_{\lambda}\sqrt{\frac{\hbar}{2\omega_{\lambda}}}\mathbf{e}_{\lambda}\left\langle m\right|b_{\lambda}^{\dagger}+b_{\lambda}\left|n\right\rangle 
\]
\[
=-\delta_{mn}\mathbf{D}_{22}^{-1}\mathbf{g}_{2}+\sum_{\lambda}\sqrt{\frac{\hbar}{2\omega_{\lambda}}}\mathbf{e}_{\lambda}\left\langle m\right|b_{\lambda}^{\dagger}+b_{\lambda}\left|n\right\rangle 
\]
Note that the matrix element in the second term is zero if $m=n$.
Hence, we obtain 
\begin{equation}
\left\langle \mathbf{x}_{2}\right\rangle _{2}^{S}=-\rho_{1}^{0}\mathbf{D}_{22}^{-1}\mathbf{g}_{2}+\sum_{m\neq n}\rho_{nm}e^{i\left(\theta_{n}-\theta_{m}\right)}\sum_{\lambda}\sqrt{\frac{\hbar}{2\omega_{\lambda}}}\mathbf{e}_{\lambda}\left\langle m\right|b_{\lambda}^{\dagger}+b_{\lambda}\left|n\right\rangle \label{eq:av-of-x2}
\end{equation}
Similarly, one has:
\begin{equation}
\left\langle \mathbf{p}_{2}\right\rangle _{2}^{S}=i\sum_{m\neq n}\rho_{nm}e^{i\left(\theta_{n}-\theta_{m}\right)}\sum_{\lambda}\sqrt{\frac{\hbar\omega_{\lambda}}{2}}\mathbf{e}_{\lambda}\left\langle m\right|b_{\lambda}^{\dagger}-b_{\lambda}\left|n\right\rangle \label{eq:p2}
\end{equation}
Substituting Eqs. (\ref{eq:av-of-x2}), (\ref{eq:p2}) and (\ref{eq:G2})
into Eq. (\ref{eq:bath-avergage-R1}), and noticing that $\dot{\mathbf{\Omega}}_{22}\left(t-t_{0}\right)\mathbf{D}_{22}^{-1}=\mathbf{\Pi}_{22}\left(t-t_{0}\right)$
\cite{my-SBC-1}, we observe that the only dependence on the states
of the open system in $\left\langle \mathbf{R}_{1}(t)\right\rangle _{2}$
comes from the operators $\rho_{mn}$. Therefore, taking the trace
over the states of the open system, one obtains the following equation
for the component $i\in1$ of the random force: 
\begin{equation}
\mathcal{R}_{i}(t)=\sum_{m\neq n}e^{i\left(\theta_{n}-\theta_{m}\right)}Z_{i}^{mn}(t)\label{eq:Random-force-c-number}
\end{equation}
where the coefficients 
\begin{equation}
Z_{i}^{mn}(t)=A_{mn}\sum_{\lambda}\left[D_{i\lambda}(t)\left\langle m\right|b_{\lambda}^{\dagger}\left|n\right\rangle +D_{i\lambda}^{*}(t)\left\langle m\right|b_{\lambda}\left|n\right\rangle \right]\label{eq:Z}
\end{equation}
do not contain the random phases. Here 
\begin{equation}
D_{i\lambda}(t)=-\sqrt{\frac{\hbar}{2\omega_{\lambda}}}V_{i\lambda}(t)e^{i\omega_{\lambda}\left(t-t_{0}\right)}\label{eq:D-coeff}
\end{equation}
and 
\begin{equation}
V_{i\lambda}(t)=\sum_{j\in2}V_{ij}(t)e_{\lambda j}\label{eq:V-1-lambda}
\end{equation}
and we have used Eqs. (\ref{eq:omega-matrix}) and (\ref{eq:omega-dot-matrix})
for the matrices $\mathbf{\Omega}_{22}$ and $\dot{\mathbf{\Omega}}_{22}$.
It is easy to see that $Z_{i}^{mn}(t)^{*}=Z_{i}^{nm}(t)$. The numbers
\begin{equation}
A_{mn}=\mbox{Tr}_{1}\left(\rho_{mn}\right)\label{eq:A_mn-via_Ro1}
\end{equation}
depend on the unknown operators $\rho_{mn}$ which act in the Hilbert
space of the open system and correspond to the initial time $t_{0}$.
Hence, in principle $A_{mn}$ would depend on the initial preparation
of the open system. 

\begin{figure}
\begin{centering}
\includegraphics[height=6cm]{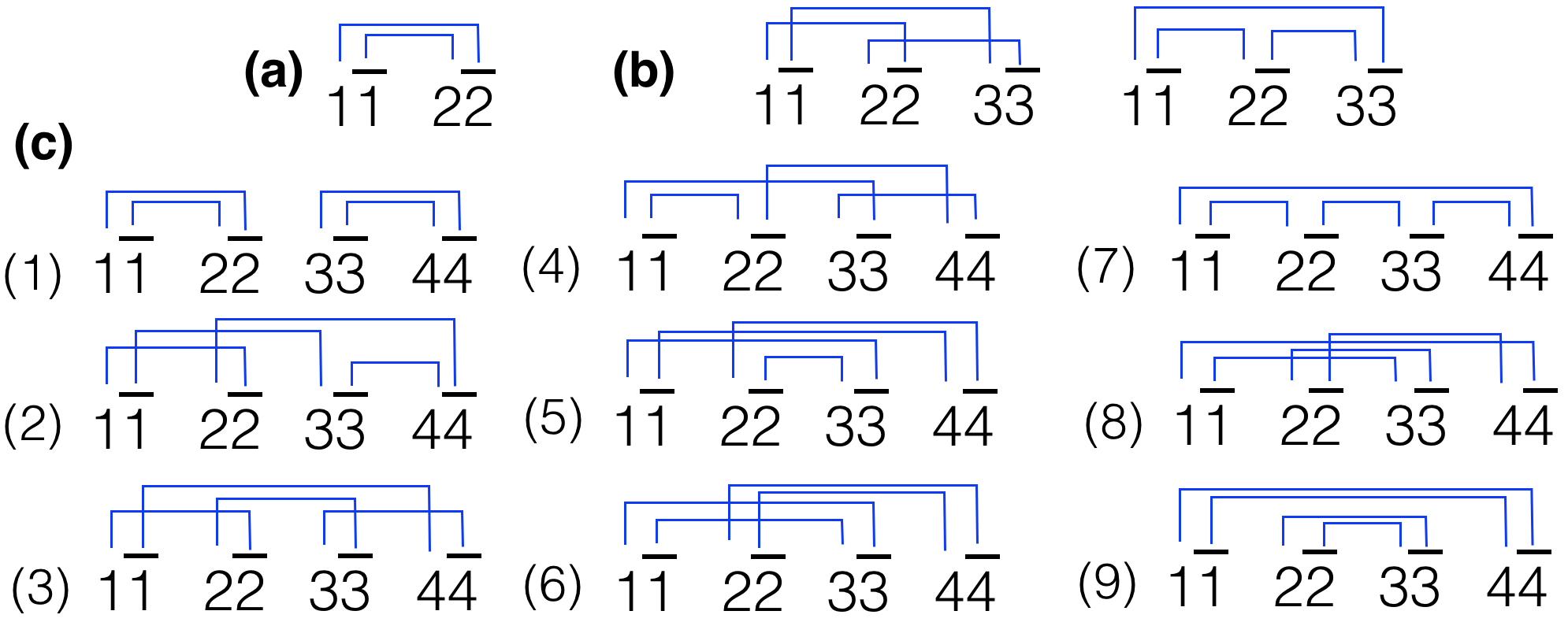}
\par\end{centering}

\caption{Schematic representation of all non-zero pairings of phases for second
(a), third (b) and fourth (c) order correlation functions. Numbers
1, 2, etc. correspond to states $n_{1}$, $n_{2}$, etc. (they have
the plus sign in the exponential factors), while a number with the
bar on top, $\overline{1}$, $\overline{2}$, etc. (they have the
minus sign in the exponentials) are associated with the states $m_{1}$,
$m_{2}$, and so on. The pairing method is somewhat similar to the
well-known Wick's theorem of the many-body quantum statistical mechanics
if numbers without the bar on top of them are associated with annihilation
operators, while the ones with the bar with creation operators; the
only difference is that pairing of the same numbers is forbidden in
our case.\label{fig:pairings}}

\end{figure}

Recall that $n$ and $m$ here represent sets of positive integers
(including zero) corresponding to quantum numbers of all vibrational
modes, i.e. $n=\left\{ n_{\lambda}\right\} $ and $m=\left\{ m_{\lambda}\right\} $.
Two such sets are considered different if at least for one mode $\lambda_{1}$
the quantum numbers differ, i.e. $n_{\lambda_{1}}\neq m_{\lambda_{1}}$.
In the following, to simplify the notations, it is convenient to ``forget''
that $n$ and $m$ are the sets of numbers and assume that they are
simply integer numbers themselves. 

Before considering the correlation functions, we note that the average
of the random force, $\overline{\mathcal{R}_{i}(t)}$, is zero since
according to Eq. (\ref{eq:average-of-exp-with-phases}) the double
sum in Eq. (\ref{eq:Random-force-c-number}) does not contain the
term $m=n$. Hence, the first moment of the random process is zero.

\subsection{Pair correlation function }

We next calculate the \emph{second order} correlation function: 
\[
\overline{\mathcal{R}_{i_{1}}\left(t_{1}\right)\mathcal{R}_{i_{2}}\left(t_{2}\right)}=\sum_{m_{1}\neq n_{1}}\:\sum_{m_{2}\neq n_{2}}\overline{e^{i\left(\theta_{n_{1}}-\theta_{m_{1}}\right)}e^{i\left(\theta_{n_{2}}-\theta_{m_{2}}\right)}}Z_{i_{1}}^{m_{1}n_{1}}\left(t_{1}\right)Z_{i_{2}}^{m_{2}n_{2}}\left(t_{2}\right)
\]
When taking the average of the exponential factors with the phases,
the integers $n_{j}$ and $m_{j}$ ($j=1,2$) may take all possible
values; they can all be different or equal, subject to the condition
that $n_{j}\neq m_{j}$ for any $j=1,2$ (see Eq. (\ref{eq:Random-force-c-number})).
Because of this condition, three or four integers cannot be equal;
a nonzero value of the average of the exponentials in the above expression
can only be possible if the four integers split into two pairs of
equal integers. Since $n_{1}\neq m_{1}$ and $n_{2}\neq m_{2}$, only
two possibilities remain which are: (i) $n_{1}=n_{2}$, $m_{1}=m_{2}$
and (ii) $n_{1}=m_{2}$, $m_{1}=n_{2}$. It is easy to see that in
the former case the contribution is zero: 
\[
\overline{e^{i\left(\theta_{n_{1}}-\theta_{m_{1}}\right)}e^{i\left(\theta_{n_{2}}-\theta_{m_{2}}\right)}}=\overline{e^{i\left(\theta_{n_{1}}-\theta_{m_{1}}\right)}e^{i\left(\theta_{n_{1}}-\theta_{m_{1}}\right)}}=\overline{e^{2i\theta_{n_{1}}}e^{-2i\theta_{m_{1}}}}=\overline{e^{2i\theta_{n_{1}}}}\,\overline{e^{-2i\theta_{m_{1}}}}=\overline{e^{2i\theta_{n_{1}}}}\,\left(\overline{e^{2i\theta_{m_{1}}}}\right)^{*}=0
\]
However, in the second case we obtain a nonzero result, 
\[
\overline{e^{i\left(\theta_{n_{1}}-\theta_{m_{1}}\right)}e^{i\left(\theta_{n_{2}}-\theta_{m_{2}}\right)}}=\overline{e^{i\left(\theta_{n_{1}}-\theta_{m_{1}}\right)}e^{i\left(\theta_{m_{1}}-\theta_{n_{1}}\right)}}=\overline{e^{i\left(\theta_{n_{1}}-\theta_{n_{1}}\right)}e^{-i\left(\theta_{m_{1}}-\theta_{m_{1}}\right)}}=\overline{1}\,\overline{1}=1
\]
yielding 
\begin{equation}
\overline{\mathcal{R}_{i_{1}}\left(t_{1}\right)\mathcal{R}_{i_{2}}\left(t_{2}\right)}=\sum_{m_{1}\neq n_{1}}\:\sum_{m_{2}\neq n_{2}}\delta_{n_{1}m_{2}}\delta_{m_{1}n_{2}}Z_{i_{1}}^{m_{1}n_{1}}\left(t_{1}\right)Z_{i_{2}}^{m_{2}n_{2}}\left(t_{2}\right)=\sum_{n_{1}n_{2}}Z_{i_{1}}^{n_{2}n_{1}}\left(t_{1}\right)Z_{i_{2}}^{n_{1}n_{2}}\left(t_{2}\right)\label{eq:R1-R1-easier}
\end{equation}
We conclude, that when pairing phases during averaging, the only non-zero
contribution came by pairing phases which have opposite signs in the
exponentials (the second case). This particular pairing can be associated
with a simple diagram shown in Fig. \ref{fig:pairings}(a). 

\begin{figure}
\begin{centering}
\includegraphics[height=2cm]{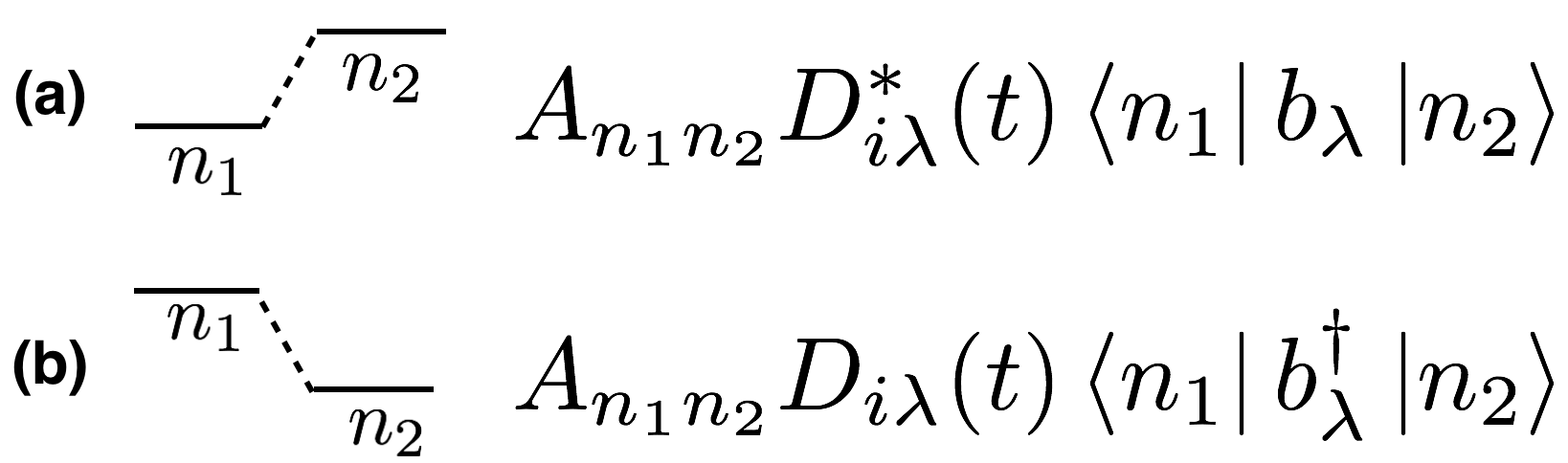}
\par\end{centering}

\caption{Two elementary graphs and the corresponding expressions associated
with them: (a) step ``up'' and (b) step ``down''. In either case
the integer numbers $n_{1}=\left\{ n_{1\lambda^{\prime}}\right\} $
and $n_{2}=\left\{ n_{2\lambda^{\prime}}\right\} $ satisfy the conditions:
$n_{1\lambda^{\prime}}=n_{2\lambda^{\prime}}$ for any $\lambda^{\prime}\protect\neq\lambda$,
and $n_{1\lambda}=n_{2\lambda}\pm1$ otherwise. \label{fig:steps}}

\end{figure}

\begin{figure}
\begin{centering}
\includegraphics[height=5cm]{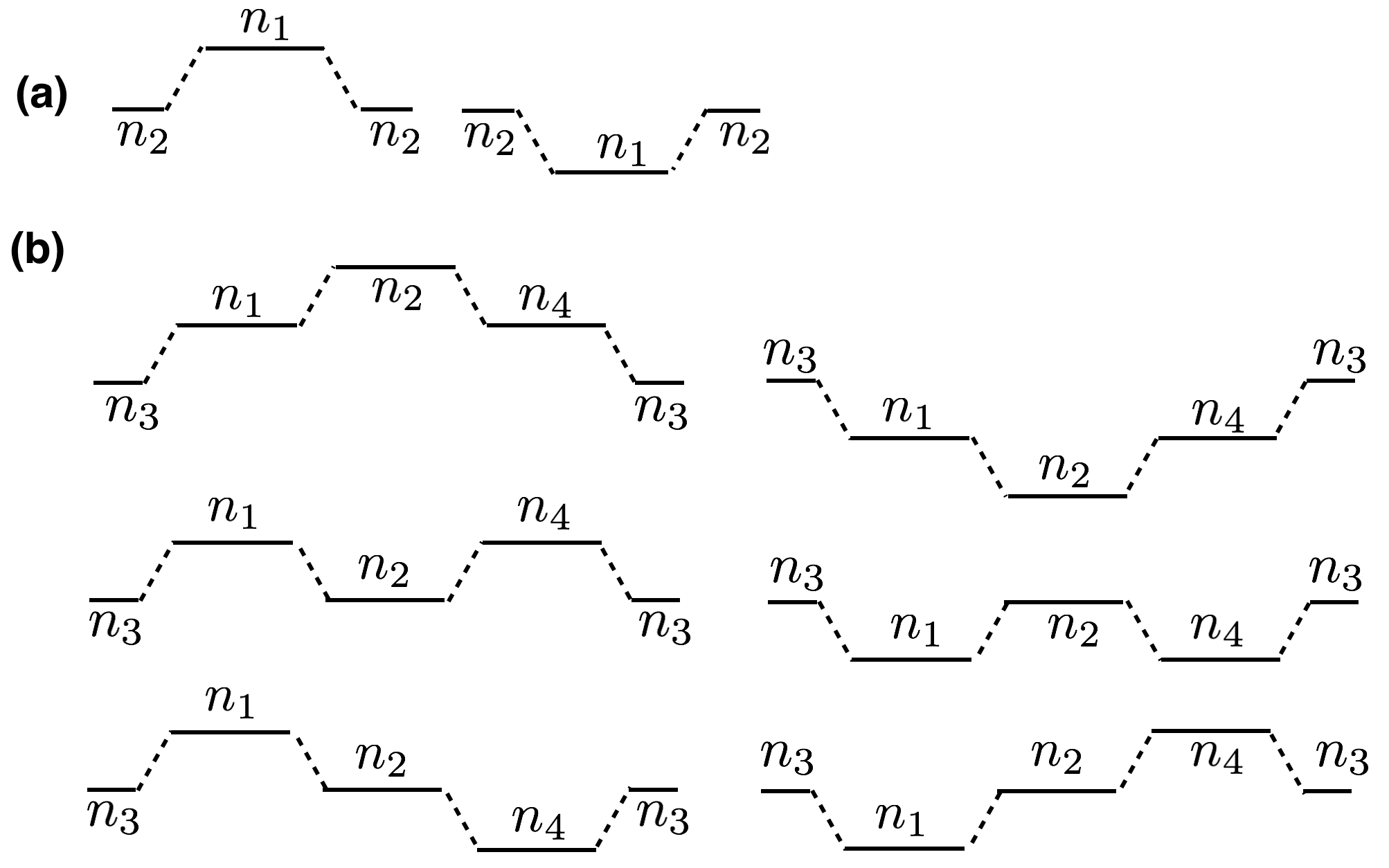}
\par\end{centering}

\caption{All step-graphs which lead to non-zero contributions for the second
order (a) and fourth order (b) correlation functions.\label{fig:step-graphs} }

\end{figure}

Next, substituting into the last expression the explicit formula (\ref{eq:Z})
for the $Z$-coefficients, we obtain:
\[
\overline{\mathcal{R}_{i_{1}}\left(t_{1}\right)\mathcal{R}_{i_{2}}\left(t_{2}\right)}=\sum_{n_{1}n_{2}}A_{n_{2}n_{1}}A_{n_{1}n_{2}}\sum_{\lambda_{1}\lambda_{2}}\left[D_{i_{1}\lambda_{1}}\left\langle n_{2}\right|b_{\lambda_{1}}^{\dagger}\left|n_{1}\right\rangle +D_{i_{1}\lambda_{1}}^{*}\left\langle n_{2}\right|b_{\lambda_{1}}\left|n_{1}\right\rangle \right]
\]
\[
\times\left[D_{i_{2}\lambda_{2}}\left\langle n_{1}\right|b_{\lambda_{2}}^{\dagger}\left|n_{2}\right\rangle +D_{i_{2}\lambda_{2}}^{*}\left\langle n_{1}\right|b_{\lambda_{2}}\left|n_{2}\right\rangle \right]
\]
where the time arguments have been omitted (they can easily be restored:
the time $t_{j}$ is placed according to $D_{i_{j}\lambda_{j}}\rightarrow D_{i_{j}\lambda_{j}}\left(t{}_{j}\right)$).
It is clear that after opening the square brackets there will be four
terms containing the following product of matrix elements:
\[
\left\langle n_{2}\right|b_{\lambda_{1}}^{\left(\dagger\right)}\left|n_{1}\right\rangle \left\langle n_{1}\right|b_{\lambda_{2}}^{\left(\dagger\right)}\left|n_{2}\right\rangle 
\]
where the dagger inside the round brackets means that the dagger either
might be there or not. Here the phonon modes $\lambda_{1}$ and $\lambda_{2}$
are independent. Clearly, no matter whether the daggers are present
or not, this product of the matrix elements can only be non-zero if
$\lambda_{1}=\lambda_{2}$. Then, we need to consider four expressions
which contain 
\[
\left\langle n_{2}\right|b_{\lambda}^{\left(\dagger\right)}\left|n_{1}\right\rangle \left\langle n_{1}\right|b_{\lambda}^{\left(\dagger\right)}\left|n_{2}\right\rangle 
\]
with the same phonon index $\lambda$. Depending on the particular
combination of daggers in the above expression, it can be either zero
or non-zero. It is convenient to associate a simple graph with each
term in the $Z$-coefficient (\ref{eq:Z}), see Fig. \ref{fig:steps}.
A non-zero contribution appears if the appropriate combination of
two elementary graphs like those shown starts and ends at the same
state $n_{2}$, see Fig. \ref{fig:step-graphs}(a). If the left graph
in Fig. \ref{fig:step-graphs}(a) results in the contribution

\[
A_{n_{2}n_{1}}D_{i_{1}\lambda}^{*}\left(t_{1}\right)\left\langle n_{2}\right|b_{\lambda}\left|n_{1}\right\rangle A_{n_{1}n_{2}}D_{i_{2}\lambda}\left(t_{2}\right)\left\langle n_{1}\right|b_{\lambda}^{\dagger}\left|n_{2}\right\rangle 
\]
\[
=\frac{\hbar}{2\omega_{\lambda}}\left|A_{n_{1}n_{2}}\left\langle n_{1}\right|b_{\lambda}^{\dagger}\left|n_{2}\right\rangle \right|^{2}V_{i_{1}\lambda}\left(t_{1}\right)V_{i_{2}\lambda}\left(t_{2}\right)e^{-i\omega_{\lambda}\left(t_{1}-t_{2}\right)}
\]
then the right one yields 
\[
A_{n_{2}n_{1}}D_{i_{1}\lambda}\left(t_{1}\right)\left\langle n_{2}\right|b_{\lambda}^{\dagger}\left|n_{1}\right\rangle A_{n_{1}n_{2}}D_{i_{2}\lambda}^{*}\left(t_{2}\right)\left\langle n_{1}\right|b_{\lambda}\left|n_{2}\right\rangle 
\]
\[
=\frac{\hbar}{2\omega_{\lambda}}\left|A_{n_{1}n_{2}}\left\langle n_{2}\right|b_{\lambda}^{\dagger}\left|n_{1}\right\rangle \right|^{2}V_{i_{1}\lambda}\left(t_{1}\right)V_{i_{2}\lambda}\left(t_{2}\right)e^{i\omega_{\lambda}\left(t_{1}-t_{2}\right)}
\]
Summing up both terms and changing the summation indices $n_{1}\longleftrightarrow n_{2}$
in the second term, we arrive at a very simple result: 
\begin{equation}
\overline{\mathbfcal{R}_{1}\left(t\right)\mathbfcal{R}_{1}^{T}\left(t^{\prime}\right)}=\mathbf{V}_{12}(t)\left[\sum_{\lambda}A_{\lambda}\mathbf{e}_{\lambda}\mathbf{e}_{\lambda}^{T}\cos\left(\omega_{\lambda}\left(t-t^{\prime}\right)\right)\right]\mathbf{V}_{21}\left(t^{\prime}\right)\label{eq:R1-R1-corr-function}
\end{equation}
By taking the transpose of both sides, it is seen that this autocorrelation
function is symmetric with respect to the permutation of times:
\begin{equation}
\left[\overline{\mathbfcal{R}_{1}\left(t\right)\mathbfcal{R}_{1}^{T}\left(t^{\prime}\right)}\right]^{T}=\overline{\mathbfcal{R}_{1}\left(t^{\prime}\right)\mathbfcal{R}_{1}^{T}\left(t\right)}\label{eq:corr-func-symmetry}
\end{equation}
Nothing can be said at this stage about the amplitudes 
\begin{equation}
A_{\lambda}=\frac{\hbar}{\omega_{\lambda}}\sum_{m\neq n}\left|A_{nm}\right|^{2}\left|\left\langle m\right|b_{\lambda}\left|n\right\rangle \right|^{2}\label{eq:A-lambda}
\end{equation}
apart from the fact that these must depend on the temperature and
the initial preparation of the system. Note also that here we basically
have a single summation over phonon states of the bath since 
\[
\left\langle m\right|b_{\lambda}\left|n\right\rangle =\left\langle \left\{ m_{\lambda^{\prime}}\right\} \right|b_{\lambda}\left|\left\{ n_{\lambda^{\prime}}\right\} \right\rangle =\sqrt{n_{\lambda}}\delta_{n_{\lambda}-1,m_{\lambda}}\left(\prod_{\lambda^{\prime}\neq\lambda}\delta_{n_{\lambda^{\prime}},m_{\lambda^{\prime}}}\right)
\]

Surprisingly, the pair correlation function (\ref{eq:R1-R1-corr-function})
has the same structure as the one in the classical GLE \cite{my-SBC-1,Lorenzo-GLE-2014}
derived for the same Hamiltonian $\mathcal{H}$, where $A_{\lambda}=1/\beta\omega_{\lambda}^{2}$.
Therefore, the unknown amplitudes $A_{\lambda}$ in Eq. (\ref{eq:R1-R1-corr-function})
are expected to tend to this limit as $\hbar,\beta\rightarrow0$.

\subsection{Odd order correlation functions}

We next consider the \emph{third order }correlation function 
\[
\overline{\mathcal{R}_{i_{1}}\left(t_{1}\right)\mathcal{R}_{i_{2}}\left(t_{2}\right)\mathcal{R}_{i_{3}}\left(t_{3}\right)}
\]
\[
=\sum_{m_{1}\neq n_{1}}\:\sum_{m_{2}\neq n_{2}}\:\sum_{m_{3}\neq n_{3}}\overline{e^{i\left(\theta_{n_{1}}-\theta_{m_{1}}\right)}e^{i\left(\theta_{n_{2}}-\theta_{m_{2}}\right)}e^{i\left(\theta_{n_{3}}-\theta_{m_{3}}\right)}}Z_{i_{1}}^{m_{1}n_{1}}\left(t_{1}\right)Z_{i_{2}}^{m_{2}n_{2}}\left(t_{2}\right)Z_{i_{3}}^{m_{3}n_{3}}\left(t_{3}\right)
\]
We ought to analyse the average of the product of the three exponentials
with the phases. In the sum above, the numbers $n_{j}$ and $m_{j}$
($j=1,2,3$) may all have different values and also there will be
identical values. Since there are limitations on the values of the
possible integers (recall that $n_{j}\neq m_{j}$ for any $j=1,2,3$),
then four, five or six integers cannot be identical. If any three
integers are identical, e.g. $n_{1}=n_{2}=m_{3}$, then the other
three must form another triple of identical numbers as well, $m_{1}=m_{2}=n_{3}$,
since otherwise the expression would contain the average of an exponential
of a single phase, $\overline{e^{i\theta}}$, which is zero. So, in
this case 

\[
\overline{e^{i\left(\theta_{n_{1}}-\theta_{m_{1}}\right)}e^{i\left(\theta_{n_{2}}-\theta_{m_{2}}\right)}e^{i\left(\theta_{n_{3}}-\theta_{m_{3}}\right)}}\:\Rightarrow\:\overline{e^{i\left(\theta_{n_{1}}-\theta_{m_{1}}\right)}e^{i\left(\theta_{n_{1}}-\theta_{m_{1}}\right)}e^{i\left(\theta_{m_{1}}-\theta_{n_{1}}\right)}}
\]
\[
=\overline{e^{i\theta_{n_{1}}}e^{-i\theta_{m_{1}}}}=\overline{e^{i\theta_{n_{1}}}}\left(\overline{e^{i\theta_{m_{1}}}}\right)^{*}=0
\]
Hence, the only possibility that remains is the one in which the six
integers are split into three \emph{pairs} of identical integers.
We have to pair up the integers of complex conjugate exponentials
(i.e. exponentials with plus and minus signs in their exponents) as
otherwise we have zero again. Indeed, if, for instance, $n_{1}=n_{2}$,
then the total average would contain a factor

\[
\overline{e^{i\theta_{n_{1}}}e^{i\theta_{n_{2}}}}\:\Rightarrow\;\overline{e^{i\theta_{n_{1}}}e^{i\theta_{n_{1}}}}=\overline{e^{2i\theta_{n_{1}}}}=0
\]
Thus, pairing only complex conjugate exponentials, we have two possibilities
shown schematically in Fig. \ref{fig:pairings}(b): (i) $n_{1}=m_{2}$,
$m_{1}=n_{3}$, $n_{2}=m_{3}$ and (ii) $n_{1}=m_{3}$, $m_{1}=n_{2}$,
$m_{2}=n_{3}$. Therefore, one can write: 
\[
\overline{e^{i\left(\theta_{n_{1}}-\theta_{m_{1}}\right)}e^{i\left(\theta_{n_{2}}-\theta_{m_{2}}\right)}e^{i\left(\theta_{n_{3}}-\theta_{m_{3}}\right)}}=
\]
\[
=\delta_{n_{1}m_{2}}\delta_{m_{1}n_{3}}\delta_{n_{2}m_{3}}\overline{e^{i\left(\theta_{n_{1}}-\theta_{m_{1}}\right)}e^{i\left(\theta_{n_{2}}-\theta_{n_{1}}\right)}e^{i\left(\theta_{m_{1}}-\theta_{n_{2}}\right)}}+\delta_{n_{1}m_{3}}\delta_{m_{1}n_{2}}\delta_{m_{2}n_{3}}\overline{e^{i\left(\theta_{n_{1}}-\theta_{m_{1}}\right)}e^{i\left(\theta_{m_{1}}-\theta_{m_{2}}\right)}e^{i\left(\theta_{m_{2}}-\theta_{n_{1}}\right)}}
\]
\[
=\delta_{n_{1}m_{2}}\delta_{m_{1}n_{3}}\delta_{n_{2}m_{3}}\overline{1}\,\overline{1}\,\overline{1}+\delta_{n_{1}m_{3}}\delta_{m_{1}n_{2}}\delta_{m_{2}n_{3}}\overline{1}\,\overline{1}\,\overline{1}=\delta_{n_{1}m_{2}}\delta_{m_{1}n_{3}}\delta_{n_{2}m_{3}}+\delta_{n_{1}m_{3}}\delta_{m_{1}n_{2}}\delta_{m_{2}n_{3}}
\]
which results in the correlation function 
\begin{equation}
\overline{\mathcal{R}_{i_{1}}\left(t_{1}\right)\mathcal{R}_{i_{2}}\left(t_{2}\right)\mathcal{R}_{i_{3}}\left(t_{3}\right)}=\sum_{n_{1}}\sum_{n_{2}}\sum_{n_{3}}\left[Z_{i_{1}}^{n_{3}n_{1}}\left(t_{1}\right)Z_{i_{2}}^{n_{1}n_{2}}\left(t_{2}\right)Z_{i_{3}}^{n_{2}n_{3}}\left(t_{3}\right)+Z_{i_{1}}^{n_{2}n_{1}}\left(t_{1}\right)Z_{i_{2}}^{n_{3}n_{2}}\left(t_{2}\right)Z_{i_{3}}^{n_{1}n_{3}}\left(t_{3}\right)\right]\label{eq:3-CF}
\end{equation}

We shall show now that either of the two terms is actually equal to
zero. As the second term is the complex conjugate of the first, it
is sufficient to consider the latter. Using the explicit expression
(\ref{eq:Z}) for the $Z$-coefficients and multiplying all terms
out, we arrive at an expression containing 8 terms altogether, each
of them being proportional to a product of three matrix elements:
\[
\left\langle n_{3}\right|b_{\lambda}^{\left(\dagger\right)}\left|n_{1}\right\rangle \left\langle n_{1}\right|b_{\lambda}^{\left(\dagger\right)}\left|n_{2}\right\rangle \left\langle n_{2}\right|b_{\lambda}^{\left(\dagger\right)}\left|n_{3}\right\rangle 
\]
where we have already set indices of all the three phonon modes to
be the same and equal to $\lambda$, as otherwise the result would
be zero. Going from left to right, in this product we start from the
state $n_{3}$ and must end at the same state. Clearly, that should
require the same number of steps ``up'' and ``down'' on the step
graphs, which is impossible with the three steps available here. Hence,
for any combination of daggers in the above expression the product
of the matrix elements is zero, leading to the zero contribution to
the third order correlation function. The same is true for the second
term in Eq. (\ref{eq:3-CF}). Thus, the third order correlation function
is equal to zero.

Similar analysis can be performed for any odd order correlation function:
the corresponding product of matrix elements will be zero as there
will only be an odd number of steps available in the step graphs.
Hence, any odd order correlation function is equal to zero.

\subsection{Even order correlation functions\label{sub:Even-order-correlation}}

Similarly one can consider higher order even correlation functions.
It follows that, when averaging over phases, that nonzero contributions
arise only by pairing integers belonging to complex conjugate exponentials
(note that integers from different pairs may coincide as well, however,
this falls within the remit of the pairing scheme, i.e. this case
does not need to be considered separately). However, we find that
the final expression does not contain a sum of products of only pair
correlation functions, see Eq. (\ref{eq:R1-R1-easier}), as there
will be cross-terms as well. 

To illustrate this point, consider the 4th order correlation function
\[
\overline{\mathcal{R}_{i_{1}}\left(t_{1}\right)\mathcal{R}_{i_{2}}\left(t_{2}\right)\mathcal{R}_{i_{3}}\left(t_{3}\right)\mathcal{R}_{i_{4}}\left(t_{4}\right)}
\]
\[
=\sum_{m_{1}\neq n_{1}}\:\sum_{m_{2}\neq n_{2}}\:\sum_{m_{3}\neq n_{3}}\:\sum_{m_{4}\neq n_{4}}\overline{e^{i\left(\theta_{n_{1}}-\theta_{m_{1}}\right)}e^{i\left(\theta_{n_{2}}-\theta_{m_{2}}\right)}e^{i\left(\theta_{n_{3}}-\theta_{m_{3}}\right)}e^{i\left(\theta_{n_{4}}-\theta_{m_{4}}\right)}}
\]
\[
\times\:Z_{i_{1}}^{m_{1}n_{1}}\left(t_{1}\right)Z_{i_{2}}^{m_{2}n_{2}}\left(t_{2}\right)Z_{i_{3}}^{m_{3}n_{3}}\left(t_{3}\right)Z_{i_{4}}^{m_{4}n_{4}}\left(t_{4}\right)
\]
When pairing the phases, nine contributions emerge overall, see Fig.
\ref{fig:pairings}(c). They are split into two types of terms. Consider
first the term associated with the pairing 1 in the Figure. It leads
to the factor $\delta_{n_{1}m_{2}}\delta_{n_{2}m_{1}}\delta_{n_{3}m_{4}}\delta_{n_{4}m_{3}}$
after averaging and hence to the following contribution to the correlation
function: 
\begin{equation}
\left[\sum_{n_{1}n_{2}}Z_{i_{1}}^{n_{1}n_{2}}\left(t_{1}\right)Z_{i_{2}}^{n_{2}n_{1}}\left(t_{2}\right)\right]\left[\sum_{n_{3}n_{4}}Z_{i_{3}}^{n_{3}n_{4}}\left(t_{3}\right)Z_{i_{4}}^{n_{4}n_{3}}\left(t_{4}\right)\right]=\overline{\mathcal{R}_{i_{1}}\left(t_{1}\right)\mathcal{R}_{i_{2}}\left(t_{2}\right)}\,\times\,\overline{\mathcal{R}_{i_{3}}\left(t_{3}\right)\mathcal{R}_{i_{4}}\left(t_{4}\right)}\label{eq:product-of-2-pairing-CFs}
\end{equation}
We see that this particular pairing scheme yields a product of two
pair correlation functions, cf. Eq. (\ref{eq:R1-R1-easier}). 

There are two more pairing schemes, 6 and 9 in Fig. \ref{fig:pairings}(c),
which lead to a product of pair correlation functions as well. This
can be seen e.g. by permuting pairs of numbers in the schemes. For
instance, after permuting $2\overline{2}\longleftrightarrow3\overline{3}$
in pairing scheme 6, it becomes identical to the pairing scheme 1.
Hence, the contribution of scheme 6 can be obtained from the above
expression (\ref{eq:product-of-2-pairing-CFs}) by the permutation
$i_{2},t_{2}\longleftrightarrow i_{3},t_{3}$. 

The second type of terms is provided by the other six pairing schemes
2, 3, 4, 5, 7, and 8 in Fig. \ref{fig:pairings}(c). Scheme 2 yields
the factor of $\delta_{n_{1}m_{2}}\delta_{n_{2}m_{4}}\delta_{n_{3}m_{1}}\delta_{n_{4}m_{3}}$
and hence the contribution 
\begin{equation}
\sum_{n_{1}n_{2}n_{3}n_{4}}Z_{i_{1}}^{n_{3}n_{1}}\left(t_{1}\right)Z_{i_{2}}^{n_{1}n_{2}}\left(t_{2}\right)Z_{i_{3}}^{n_{4}n_{3}}\left(t_{3}\right)Z_{i_{4}}^{n_{2}n_{4}}\left(t_{4}\right)\label{eq:nasty-contribut}
\end{equation}
to the correlation function. This term cannot be split into a simple
product of two pair correlation functions since all four $Z$-coefficients
are coupled. Other pairing schemes 3, 4, 5, 7 and 8 can be related
to this one by an appropriate permutation of the pairs of numbers.
For instance, pairing scheme 3 is brought into 2 by the permutation
$4\overline{4}\longleftrightarrow3\overline{3}$, and hence the final
contribution of pairing 3 is obtained from Eq. (\ref{eq:nasty-contribut})
by the permutation $i_{4},t_{4}\longleftrightarrow i_{3},t_{3}$.
The same can be done for all other pairing schemes belonging to this
second type of pairings. 

Therefore, it is sufficient to consider just one pairing scheme of
this type. We shall consider the scheme 2. First, we reorder terms
in the product (\ref{eq:nasty-contribut}) of the $Z$-coefficients:
\[
Z_{i_{1}}^{n_{3}n_{1}}\left(t_{1}\right)Z_{i_{2}}^{n_{1}n_{2}}\left(t_{2}\right)Z_{i_{4}}^{n_{2}n_{4}}\left(t_{4}\right)Z_{i_{3}}^{n_{4}n_{3}}\left(t_{3}\right)
\]
This ensures the continuous flow of the state numbers, $n_{3}\rightarrow n_{1}\rightarrow n_{2}\rightarrow n_{4}\rightarrow n_{3}$,
when reading from left to right. Hence, we can now turn to step graphs
which start and end at the same state $n_{3}$. There are four $Z$-coefficients
and hence we can make four steps overall; 2 steps up and 2 steps down
would give nonzero contributions. There are 6 such step graphs possible,
all shown in Fig. \ref{fig:step-graphs}(b), any of them gives a nonzero
contribution. The graphs in the Figure are arranged in pairs horizontally
which provide contributions that are complex conjugate to each other.
The top left graph in Fig. \ref{fig:step-graphs}(b) is associated
with the contribution
\[
\sum_{\lambda}\sum_{n_{1}n_{2}n_{3}n_{4}}\delta_{n_{1}n_{4}}A_{n_{3}n_{1}}A_{n_{1}n_{2}}A_{n_{2}n_{1}}A_{n_{1}n_{3}}D_{i_{1}\lambda}^{*}D_{i_{2}\lambda}^{*}D_{i_{3}\lambda}D_{i_{4}\lambda}\left|\left\langle n_{1}\right|b_{\lambda}^{\dagger}\left|n_{3}\right\rangle \left\langle n_{2}\right|b_{\lambda}^{\dagger}\left|n_{1}\right\rangle \right|^{2}
\]
\[
=\sum_{\lambda}A_{\lambda}^{(a)}V_{i_{1}\lambda}\left(t_{1}\right)V_{i_{2}\lambda}\left(t_{2}\right)V_{i_{3}\lambda}\left(t_{3}\right)V_{i_{4}\lambda}\left(t_{4}\right)e^{-i\omega_{\lambda}\left(t_{1}+t_{2}-t_{3}-t_{4}\right)}
\]
where
\[
A_{\lambda}^{(a)}=\left(\frac{\hbar}{2\omega_{\lambda}}\right)^{2}\sum_{n_{1}n_{2}n_{3}}\left|A_{n_{3}n_{1}}A_{n_{1}n_{2}}\left\langle n_{1}\right|b_{\lambda}^{\dagger}\left|n_{3}\right\rangle \left\langle n_{2}\right|b_{\lambda}^{\dagger}\left|n_{1}\right\rangle \right|^{2}
\]
is a positive real factor. Note that because of the matrix elements,
the states in the triple sum above are constrained by the conditions:
$n_{1}\neq n_{2},n_{3}$. 

The step graph on the right of the one we have just considered results
in a complex conjugate contribution, leading therefore to the following
real contribution from both these graphs: 
\begin{equation}
2\sum_{\lambda}A_{\lambda}^{(a)}V_{i_{1}\lambda}\left(t_{1}\right)V_{i_{2}\lambda}\left(t_{2}\right)V_{i_{3}\lambda}\left(t_{3}\right)V_{i_{4}\lambda}\left(t_{4}\right)\cos\left(\omega_{\lambda}\left(t_{1}+t_{2}-t_{3}-t_{4}\right)\right)\label{eq:4-term1}
\end{equation}

Similarly the other two pairs of the step graphs can be considered,
resulting in the following contribuitons:
\begin{equation}
2\sum_{\lambda}A_{\lambda}^{(b)}V_{i_{1}\lambda}\left(t_{1}\right)V_{i_{2}\lambda}\left(t_{2}\right)V_{i_{3}\lambda}\left(t_{3}\right)V_{i_{4}\lambda}\left(t_{4}\right)\cos\left(\omega_{\lambda}\left(t_{1}-t_{2}-t_{3}+t_{4}\right)\right)\label{eq:4-term2}
\end{equation}
is associated with the two graphs in the middle of Fig. \ref{fig:step-graphs}(b),
while 
\begin{equation}
2\sum_{\lambda}A_{\lambda}^{(a)}V_{i_{1}\lambda}\left(t_{1}\right)V_{i_{2}\lambda}\left(t_{2}\right)V_{i_{3}\lambda}\left(t_{3}\right)V_{i_{4}\lambda}\left(t_{4}\right)\cos\left(\omega_{\lambda}\left(t_{1}-t_{2}+t_{3}-t_{4}\right)\right)\label{eq:4-term3}
\end{equation}
is attributed to the two graphs at the bottom. Here
\[
A_{\lambda}^{(b)}=\left(\frac{\hbar}{2\omega_{\lambda}}\right)^{2}\sum_{n_{1}n_{2}}\left|A_{n_{1}n_{2}}\left\langle n_{2}\right|b_{\lambda}^{\dagger}\left|n_{1}\right\rangle \right|^{4}
\]
The sum of the three contributions (\ref{eq:4-term1})-(\ref{eq:4-term3})
corresponds to pairing scheme 2 in Fig. \ref{fig:pairings}(c). 

Other pairing schemes 3, 4, 5, 7 and 8 in the Figure are obtained
by permuting indices $i_{j}$ and the times $t_{j}$ as was explained
above. As can be seen from the formulae (\ref{eq:4-term1})-(\ref{eq:4-term3}),
each contribution is symmetric with respect to the permutations of
the indices $i_{j}$, so only times need to be permuted. The final
expression for the correlation function, containing contributions
from all pairing schemes, becomes fully symmetric with respect to
permutations of times. The correlation function depends on two unknown
coefficients $A_{\lambda}^{(a)}$ and $A_{\lambda}^{(b)}$ for each
normal mode $\lambda$ of the bath. 

Hence, the 4th order correlation function, apart from the three terms
corresponding to a product of all possible pair correlation functions,
contains additional non-zero terms which cannot be represented as
a product of pair correlations functions. 

The method developed above can be applied without difficulty to higher
(even) order correlation functions if necessary.

\subsection{Are the stochastic forces Gaussian?}

We conclude that the higher order correlation functions do not fully
satisfy the properties of a Gaussian stochastic process: although
correlation functions of any odd order are zero, even order correlation
functions do not split solely into a sum of products of pair correlation
functions; there are additional nonzero terms as well. This property
of the stochastic forces poses a certain difficulty in numerical simulations
as, at variance with the Gaussian stochastic forces of the classical
case, a two-force correlation function is not sufficient for numerical
simulations in the quantum case, i.e. higher order correlation functions
need also be considered when generating the stochastic forces in actual
numerical simulations. As the simplest approximation, one can assume
that the stochastic forces in the $c-$number quantum GLE equations
are Gaussian in which case only the lowest order (pair) correlation
function suffices.

\section{Sampling over the stochastic field}

Because of the way the stochastic field has been introduced, exact
results can only formally be obtained by averaging the calculated
trajectories using different sampling of the stochastic field. This
follows from the fact that we obtain the same equations of motion
for the exact mean values $\left\langle \mathbf{x}_{1}\right\rangle _{t}=\mbox{Tr}\left(\rho^{0}\widetilde{\mathbf{x}}_{1}(t)\right)$
after averaging over the stochastic field and after calculating the
expectation values of the position operator $\mathbf{x}_{1}$ without
the stochastic field in it, i.e. using directly the density matrix
(\ref{eq:Ro_0}). 

Indeed, let us first perform the stochastic averaging of the equations
of motion (\ref{eq:QGLE-1}). Since $\overline{\rho^{S}}=\rho^{0}$,
the averaged mean values $\overline{\left\langle \mathbf{x}_{1}\right\rangle _{t}^{S}}=\mbox{Tr}\left(\overline{\rho^{S}}\widetilde{\mathbf{x}}_{1}(t)\right)$
can be replaced with the exact ones, $\left\langle \mathbf{x}_{1}\right\rangle _{t}=\mbox{Tr}\left(\rho^{0}\widetilde{\mathbf{x}}_{1}(t)\right)$.
Next, since the stochastic average of the random force is zero, the
equation for the mean values of the positions reads:
\begin{equation}
\mathbf{M}_{11}\partial_{t}^{2}\left\langle \mathbf{x}_{1}\right\rangle _{t}=\mathcal{\mathbf{F}}_{1}(t)-\int_{t_{0}}^{t}\mathbf{K}_{11}\left(t,\tau\right)\partial_{t}\left\langle \mathbf{x}_{1}\right\rangle _{\tau}d\tau\label{eq:QGLE-1-1}
\end{equation}

It is easily checked that the same equation is obtained directly by
performing quantum statistical average of the Heisenberg equations
(\ref{eq:for-X1}) using the exact density matrix $\rho^{0}$ of Eq.
(\ref{eq:Ro_0}). To this end, we multiply both sides of this equation
by $\rho^{0}$ and take the trace over the whole system. The average
of the random force operator $\mathbf{R}_{1}(t)$ reads 
\[
\left\langle \mathbf{R}_{1}(t)\right\rangle _{t}=\mbox{Tr}_{1}\left\langle \mathbf{R}_{1}(t)\right\rangle _{2}=-\mathbf{V}_{12}(t)\left[\dot{\mathbf{\Omega}}_{22}\left(t-t_{0}\right)\left\langle \mathbf{x}_{2}\right\rangle _{t}+\mathbf{\Omega}_{22}\left(t-t_{0}\right)\left\langle \mathbf{p}_{2}\right\rangle _{t}+\mathbf{\Pi}_{22}\left(t-t_{0}\right)\left\langle \mathbf{g}_{2}\right\rangle _{t}\right]
\]
where $\left\langle \mathbf{g}_{2}\right\rangle _{t}=\mathbf{V}_{2}\left(t_{0}\right)$
(see the definition (\ref{eq:g2})). Using Eq. (\ref{eq:Ro_0}), we
write: 
\[
\left\langle \mathbf{x}_{2}\right\rangle _{t}=\mbox{Tr}_{1}\sum_{n}\rho_{n}^{eq}\rho_{1}^{0}\left\langle \psi_{n}\right|\mathbf{x}_{2}\left|\psi_{n}\right\rangle =-\mathbf{D}_{22}^{-1}\mbox{Tr}_{1}\left(\rho_{1}^{0}\mathbf{g}_{2}\right)=-\mathbf{D}_{22}^{-1}\left\langle \mathbf{g}_{2}\right\rangle _{t}
\]
and $\left\langle \mathbf{p}_{2}\right\rangle _{t}=\mathbf{0}$, cf.
the derivation of Eqs. (\ref{eq:av-of-x2}) and (\ref{eq:p2}), and
we obtain $\left\langle \mathbf{R}_{1}(t)\right\rangle _{t}=\mathbf{0}$.
Hence, multiplying both sides of Eq. (\ref{eq:for-X1}) by $\rho^{0}$,
taking the trace over the whole system and using the fact that $\mbox{Tr}\left(\rho^{0}\widetilde{\mathbf{u}}_{1}\right)=\mathbf{0}$,
we arrive at the same Eq. (\ref{eq:QGLE-1-1}) for the exact mean
values. 

We conclude that exact results can only formally be obtained by averaging
the calculated trajectories over different realisations of the stochastic
field.

\section{The closure relationship}

Until now we have demonstrated that there exists a class of cQGLEs
of the form similar to the classical GLE and containing a generally
non-Gaussian random forces. If one assumes that the forces are approximately
Gaussian, then the pair correlation function is sufficient to generate
them. However, in order to know the pair correlation function, according
to Eq. (\ref{eq:R1-R1-corr-function}), one need to determine the
unknown amplitudes $A_{\lambda}$. Here we shall propose a heuristic
argument that yields an explicit expression for $A_{\lambda}$ with
the correct classical limit. It is based on an assumption that the
correlation function of the random force does not depend on the order
in which the averages are taken. 

Eq. (\ref{eq:R1-R1-corr-function}) was derived with the trace over
both regions (the quantum statistical average) taken first to yield
$\mathbfcal{R}_{1}(t)=\mbox{Tr}\left(\mathbf{R}_{1}(t)\right)$, and
only after that the average over the stochastic field was applied
when calculating the correlation function over the stochastic field
of the phases. However, the calculation can also be done in a different
order: first, we average with respect to the stochastic field (which
results in the initial density matrix since $\overline{\rho^{S}}\equiv\rho^{0}$),
and only then the trace with the averaged density matrix $\rho^{0}$
is performed. Our hypothesis states that the correlation functions
calculated in both ways must coincide. Since the correlation function
(\ref{eq:R1-R1-corr-function}) is symmetric with respect to the time
permutation, $t\leftrightarrow t^{\prime}$ (and is real), the required
condition (which we shall call the \emph{closure relation}ship) is:

\begin{equation}
\overline{\mathbfcal{R}_{1}\left(t\right)\mathbfcal{R}_{1}^{T}\left(t^{\prime}\right)}=\frac{1}{2}\left[\left\langle \mathbf{R}_{1}\left(t\right)\mathbf{R}_{1}^{\dagger}\left(t^{\prime}\right)\right\rangle _{av}+\left\langle \mathbf{R}_{1}\left(t^{\prime}\right)\mathbf{R}_{1}^{\dagger}\left(t\right)\right\rangle _{av}\right]\label{eq:closure-Def}
\end{equation}
Here the operator $\mathbf{R}_{1}(t)$ is defined by Eq. (\ref{eq:random-force1}).
In calculating the averages $\left\langle \ldots\right\rangle _{av}=\mbox{Tr}_{1}\mbox{Tr}_{2}\left(\rho_{0}\ldots\right)=\mbox{Tr}_{1}\left\langle \ldots\right\rangle _{2}$,
we use the explicit expression (\ref{eq:Ro_0}) for $\rho^{0}$. 

When calculating the whole trace $\left\langle \ldots\right\rangle _{av}$,
it is convenient to perform the trace over the bath first:

\[
\left\langle \mathbf{R}_{1}\left(t\right)\mathbf{R}_{1}^{\dagger}\left(t^{\prime}\right)\right\rangle _{2}=\rho_{1}^{0}\sum_{m}\rho_{m}^{eq}\left\langle \psi_{m}\right|\mathbf{R}_{1}\left(t\right)\mathbf{R}_{1}^{\dagger}\left(t^{\prime}\right)\left|\psi_{m}\right\rangle 
\]
\[
=\rho_{1}^{0}\sum_{m}\rho_{m}^{eq}\left\langle m\right|\mathcal{U}^{\dagger}\mathbf{R}_{1}\left(t\right)\mathbf{R}_{1}^{\dagger}\left(t^{\prime}\right)\mathcal{U}\left|m\right\rangle =\rho_{1}^{0}\sum_{m}\rho_{m}^{eq}\left\langle m\right|\widetilde{\mathbf{R}}_{1}(t)\widetilde{\mathbf{R}}_{1}^{T}\left(t^{\prime}\right)\left|m\right\rangle 
\]
where $\widetilde{\mathbf{R}}_{1}(t)=\mathcal{U}^{\dagger}\mathbf{R}_{1}(t)\mathcal{U}$.
To calculate the latter force operator, it is useful first to simplify
the expression for the force. From Eq. (\ref{eq:random-force1})
\[
\mathbf{R}_{1}(t)=-\sum_{\lambda}\sqrt{\frac{\hbar}{2\omega_{\lambda}}}\mathbf{V}_{12}(t)\left[\left(\dot{\mathbf{\Omega}}_{22}\left(t-t_{0}\right)+i\omega_{\lambda}\mathbf{\Omega}_{22}\left(t-t_{0}\right)\right)e_{\lambda}b_{\lambda}^{\dagger}+\left(\dot{\mathbf{\Omega}}_{22}\left(t-t_{0}\right)-i\omega_{\lambda}\mathbf{\Omega}_{22}\left(t-t_{0}\right)\right)e_{\lambda}b_{\lambda}\right]
\]
\[
-\mathbf{V}_{12}(t)\mathbf{\Pi}_{22}\left(t-t_{0}\right)\mathbf{g}_{2}
\]
Using Eqs. (\ref{eq:omega-matrix}) and (\ref{eq:omega-dot-matrix}),
we find that 
\[
\left(\dot{\mathbf{\Omega}}_{22}\left(t-t_{0}\right)+i\omega_{\lambda}\mathbf{\Omega}_{22}\left(t-t_{0}\right)\right)\mathbf{e}_{\lambda}=\mathbf{e}_{\lambda}e^{i\omega_{\lambda}\left(t-t_{0}\right)}
\]
so that we finally obtain for the force operator an expression: 

\begin{equation}
\mathbf{R}_{1}(t)=\sum_{\lambda}\left[\mathbf{D}_{1\lambda}(t)b_{\lambda}^{\dagger}+\mathbf{D}_{1\lambda}^{*}(t)b_{\lambda}\right]-\mathbf{V}_{12}(t)\mathbf{\Pi}_{22}\left(t-t_{0}\right)\mathbf{g}_{2}\label{eq:R1-simplier}
\end{equation}
where the elements of the vector $\mathbf{D}_{1\lambda}(t)=\left(D_{i\lambda},i\in1\right)$
are given by Eq. (\ref{eq:D-coeff}).

Then using the fact that $\mathcal{U}^{\dagger}b_{\lambda}\mathcal{U}=b_{\lambda}+\gamma_{\lambda}$
and $\mathcal{U}^{\dagger}b_{\lambda}^{\dagger}\mathcal{U}=b_{\lambda}^{\dagger}+\gamma_{\lambda}$,
we easily obtain: 
\begin{equation}
\widetilde{\mathbf{R}}_{1}(t)=\sum_{\lambda}\left[\mathbf{D}_{1\lambda}(t)b_{\lambda}^{\dagger}+\mathbf{D}_{1\lambda}^{*}(t)b_{\lambda}\right]\label{eq:R1-simplier-1}
\end{equation}
We see that the third term in Eq. (\ref{eq:R1-simplier}) disappears
completely, which renders the tilde-force operator $\widetilde{\mathbf{R}}_{1}(t)$
to be independent of the displacement vector $\mathbf{u}_{1}$. Hence,
the trace over the open system is trivially calculated resulting in
the following expression for the correlation function: 
\[
\left\langle \mathbf{R}_{1}\left(t\right)\mathbf{R}_{1}^{\dagger}\left(t^{\prime}\right)\right\rangle _{av}=\sum_{m}\rho_{m}^{eq}\left\langle m\right|\widetilde{\mathbf{R}}_{1}(t)\widetilde{\mathbf{R}}_{1}^{T}\left(t^{\prime}\right)\left|m\right\rangle =\left\langle \widetilde{\mathbf{R}}_{1}(t)\widetilde{\mathbf{R}}_{1}^{T}\left(t^{\prime}\right)\right\rangle _{2}^{eq}
\]
This is nothing but the trace over the bath using its equilibrium
density matrix. Therefore, from Eq. (\ref{eq:R1-simplier-1}) it is
clear that, upon multiplication of the tilda-forces, only products
of annihilation and creation operators will contribute:
\begin{equation}
\left\langle \widetilde{\mathbf{R}}_{1}(t)\widetilde{\mathbf{R}}_{1}^{T}\left(t^{\prime}\right)\right\rangle _{2}^{eq}=\sum_{\lambda\lambda^{\prime}}\left\{ \left[\mathbf{D}_{1\lambda}(t)\mathbf{D}_{1\lambda^{\prime}}^{\dagger}\left(t^{\prime}\right)\right]\left\langle b_{\lambda}^{\dagger}b_{\lambda^{\prime}}\right\rangle _{2}^{eq}+\left[\mathbf{D}_{1\lambda}(t)\mathbf{D}_{1\lambda^{\prime}}^{\dagger}\left(t^{\prime}\right)\right]^{*}\left\langle b_{\lambda}b_{\lambda^{\prime}}^{\dagger}\right\rangle _{2}^{eq}\right\} \label{eq:pair-corr}
\end{equation}
Since 
\[
\mathbf{D}_{1\lambda}(t)\mathbf{D}_{1\lambda^{\prime}}^{\dagger}\left(t^{\prime}\right)=\mathbf{V}_{12}\left(t\right)\left[\frac{\hbar}{2\omega_{\lambda}}\mathbf{e}_{\lambda}\mathbf{e}_{\lambda}^{T}e^{i\omega_{\lambda}\left(t-t^{\prime}\right)}\right]\mathbf{V}_{21}\left(t^{\prime}\right)
\]
and $\left\langle b_{\lambda}^{\dagger}b_{\lambda^{\prime}}\right\rangle _{2}^{eq}=\left\langle b_{\lambda}b_{\lambda^{\prime}}^{\dagger}\right\rangle _{2}^{eq}-\delta_{\lambda\lambda^{\prime}}=\delta_{\lambda\lambda^{\prime}}\overline{n}_{\lambda}=\delta_{\lambda\lambda^{\prime}}\left(e^{\beta\hbar\omega_{\lambda}}-1\right)^{-1}$,
we obtain:

\begin{equation}
\left\langle \mathbf{R}_{1}\left(t\right)\mathbf{R}_{1}^{\dagger}\left(t^{\prime}\right)\right\rangle _{av}=\mathbf{V}_{12}(t)\left[\sum_{\lambda}\frac{\hbar}{\omega_{\lambda}}\left(\overline{n}_{\lambda}+\frac{1}{2}\right)\mathbf{e}_{\lambda}\mathbf{e}_{\lambda}^{T}\cos\left(\omega_{\lambda}\left(t-t^{\prime}\right)\right)-\frac{i\hbar}{2}\mathbf{\Omega}_{22}\left(t-t^{\prime}\right)\right]\mathbf{V}_{21}\left(t^{\prime}\right)\label{eq:R-R-corr-func}
\end{equation}
where Eq. (\ref{eq:omega-matrix}) was also used. Note that the second
term is purely imaginary and is antisymmetric with respect to the
time permutation. This term disappears when using the above expression
in the closure relationship, Eq. (\ref{eq:closure-Def}), which yields:
\begin{equation}
\overline{\mathbfcal{R}_{1}\left(t\right)\mathbfcal{R}_{1}^{T}\left(t^{\prime}\right)}=\mathbf{V}_{12}(t)\left[\sum_{\lambda}\frac{\hbar}{\omega_{\lambda}}\left(\overline{n}_{\lambda}+\frac{1}{2}\right)\mathbf{e}_{\lambda}\mathbf{e}_{\lambda}^{T}\cos\left(\omega_{\lambda}\left(t-t^{\prime}\right)\right)\right]\mathbf{V}_{21}\left(t^{\prime}\right)\label{eq:R1-R1-after-closure}
\end{equation}
The expression on the right hand side has the same form as expression
(\ref{eq:R1-R1-corr-function}), where we used a different ordering
for the averages (i.e. first the quantum statistical average and then
the stochastic average). Comparing these two expressions, we obtain
the following formula for the amplitudes we have been looking for:
\begin{equation}
A_{\lambda}=\frac{\hbar}{\omega_{\lambda}}\left(\overline{n}_{\lambda}+\frac{1}{2}\right)=\frac{\hbar}{2\omega_{\lambda}}\coth\left(\frac{1}{2}\beta\hbar\omega_{\lambda}\right)\label{eq:amplitudes-final}
\end{equation}
It can be seen that this expression for the amplitude tends to the
correct classical limit, $A_{\lambda}\rightarrow1/\beta\omega_{\lambda}^{2}$,
when either $\hbar\rightarrow0$ or $\beta\rightarrow0$. It is highly
encouraging that the same form of the pair correlation function has
been obtained by the two methods. Moreover, the second method enabled
us to \emph{propose }an explicit expression for the previously unknown
amplitudes $A_{\lambda}$. The result we obtained basically coincides
with the expression from \cite{Benguria-Kac-1981,FKM-J.Math.Ph.-1965,Ford-Kac-JST-1987,Gardiner-book_noise-2010}
for the correlation function that was used in \cite{Banerjee-Bag-Banik-Roy-PRE-2002,Banerjee-Banik-Bag-Ray-PHE-2002,Banik-Bag-Ray-PRE-2002,Banerjee-Bag-Banik-Ray-JCP-2004}
without a proper justification (and for a more simplified Hamiltonian).

Unfortunately, an analogous procedure does not seem to exist for higher
order correlation functions. This is because an average of a product
of the force correlations functions, 
\begin{equation}
\left\langle R_{i_{1}}\left(t_{1}\right)R_{i_{2}}\left(t_{2}\right)R_{i_{3}}\left(t_{3}\right)\ldots\right\rangle _{av}=\left\langle \widetilde{R}_{i_{1}}\left(t_{1}\right)\widetilde{R}_{i_{2}}\left(t_{2}\right)\widetilde{R}_{i_{3}}\left(t_{3}\right)\ldots\right\rangle _{2}^{eq}\label{eq:averages}
\end{equation}
is given by the quantum-statistical average over the equilibrium bath,
as given above, and hence the Wick's theorem can be used to calculate
it. An appropriate calculation shows that odd order averages are all
zero, but even order ones split into a sum of products of only pair
correlations (\ref{eq:pair-corr}). Hence, the extra terms we encountered
in Section \ref{sub:Even-order-correlation} do not appear at all
in the averages (\ref{eq:averages}). So the question of obtaining
unknown amplitudes in the extra terms in the even order correlation
functions (as e.g. $A_{\lambda}^{(a)}$ and $A_{\lambda}^{(b)}$ from
Section \ref{sub:Even-order-correlation}), still remains open. One
must expect that these extra terms that cannot be represented as a
product of pair correlation functions, must tend to zero in the classical
limit of $\hbar,\beta\rightarrow0$. This would ensure that in the
classical limit the random forces correspond to a Gaussian stochastic
process.

\section{Conclusions}

Concluding, in this paper we offer a derivation of a fully quantum
$c-$number GLE which is a self-contained equation for the expectation
values of the positions of atoms in the open system. Our method is
based on a rather general Hamiltonian of the combined system (the
open system and bath) which possesses a surprising similarity with
its classical analogue \cite{my-SBC-1}: an identical friction kernel
and a similar structure of the random force autocorrelation function.
Our derivation is based on the harmonisation approximation whereby
a harmonic expansion is made around the exact\emph{ instantaneous
}mean values of the positions of atoms in the open system which \emph{evolve
in time}. A possible direction for future research is to go beyond
this approximation e.g. along the lines proposed in \cite{McDowell-JCP-2000,Banerjee-Banik-Bag-Ray-PHE-2002}.
No product approximation was assumed in our treatment for the initial
density matrix of the combined system as done e.g. in the Feynman-Vernon
method \cite{Feynman-Vernon-1963,Kleinert-Shabanov-PLA-1995} and
some of the traditional quantum (operator) GLE approaches \cite{Gardiner-book_noise-2010}.
Our equations contain non-Gaussian stochastic forces, which have zero
mean and appropriate correlation functions, and perform as a coloured
noise. The simplest pair correlation function has the same functional
form as in the classical case \cite{my-SBC-1}, but contains amplitudes
$A_{\lambda}$ which may depend on the initial preparation of the
open system and, of course, on the temperature.

It is also shown that after sampling over many trajectories due to
different realisations of the stochastic field, our approach converges
to the exact trajectory for the mean values $\left\langle x_{1}\right\rangle _{t}$
of the open system atomic positions.

In the first approximation, stochastic forces may be considered as
Gaussian. To offer a practical computational scheme, an explicit expression
for the amplitudes of the pair correlation function was offered. It
is based on a conjecture that no matter in which order the stochastic
and statistical averages are taken when calculating the random force
autocorrelation function, the same result is to be expected. The obtained
expression for the amplitudes has the correct classical limit.

Our method sets a foundation for a practical "classical"-like computational technique,
which could be used for calculating atomic trajectories in an open
system under arbitrary non-equilibrium conditions - these would be
fully quantum MD simulations for mean atomic positions. Note that
quantum MD simulations based on path integrals \cite{Habershon-ARPC-2013,Herrero-Ramirez-JPCM-2014}
are designed only for thermodynamic equilibrium.

One may ask why perform quantum MD simulations with stochastic forces
with subsequent averaging over many realisations of the stochastic
field instead of solving directly the equations for the mean atomic
positions that do not contain the stochastic force? There are at least
two advantages in using stochastic methods: (i) stochastic equations
of motion give access to fluctuations of atomic trajectories from
the mean trajectory, and (ii) there exist powerful numerical techniques
for solving stochastic differential equations with the memory and random forces \cite{Lorenzo-GLE-2014,Herve-GLE-2015,Herve-GLE-2016}
which can be exploited (assuming that stochastic forces are Gaussian). Of course,
further work is needed in designing a computational scheme which accounts for
the non-Gaussian character of the random force.

\section*{Acknowledgements}
We acknowledge financial support from the UK EPSRC under Grant No.
EP/J019259/1. HN, CDL and LK acknowledge the stimulating research
environment provided by the EPSRC Centre for Doctoral Training in
Cross-Disciplinary Approaches to Non-Equilibrium Systems (CANES, EP/L015854/1).
We would also like to acknowledge a financial support of the COST action MP1303 "Understanding and controlling nano and mesoscale friction" for facilitating valuable discussions.


\begin{thebibliography}{10}

\bibitem{Petuccione-Open-Systems-2007}
H.-P. Breuer and F.~Petruccione,
\newblock {\em The Theory of Open Quantum Systems},
\newblock Oxford Univ. Press, 2007.

\bibitem{Weiss-2012}
U.~Weiss,
\newblock {\em Quantum Dissipative System},
\newblock World Scientific, Singapore, 2012.

\bibitem{Szlufarska2007}
I. Szlufarska, R.~K. Kalia, A. Nakano, and P. Vashishta,
\newblock {\em A molecular dynamics study of nanoindentation of amorphous silicon carbide},
\newblock J. Appl. Phys. {\bf 102}, 023509 (2007).

\bibitem{Barry2009}
P.~R. Barry, P.~Y. Chiu, S.~S. Perry, W.~G. Sawyer, S.~R. Phillpot, and S.~B.
  Sinnott,
\newblock {\em The effect of normal load on polytetrafluoroethylene tribology},
\newblock J.Phys.: Condens. Matter {\bf 21}, 144201 (2009).

\bibitem{Lorenz2010}
Christian~D. Lorenz, Michael Chandross, and Gary~S. Grest,
\newblock {\em Large scale molecular dynamics simulations of vapor phase lubrication
  for MEMS},
\newblock J. Adh. Sci. Tech. {\bf   24}, 2453 (2010).

  
\bibitem{Trevethan2004}
T.~Trevethan and L.~Kantorovich,
\newblock {\em Stochastic mechanism of energy dissipation in noncontact atomic force
  microscopy studied using molecular dynamics with Langevin boundary
  conditions},
\newblock Phys. Rev. B, {\bf 70}, 115411 (2004).

\bibitem{Kermode2008}
J.~R. Kermode, T.~Albaret, D.~Sherman, N.~Bernstein, P.~Gumbsch, M.~C. Payne,
  G.~Cs\'{a}nyi, and A.~De~Vita,
\newblock {\em Low-speed fracture instabilities in a brittle crystal},
\newblock Nature {\bf 455}, 1224 (2008).

\bibitem{Mazyar2006}
Oleg~A. Mazyar and William~L. Hase,
\newblock {\em Dynamics and kinetics of heat transfer at the interface of model
  diamond {111} nanosurfaces},
\newblock J. Phys. Chem. A {\bf 110}, 526 (2006).

\bibitem{Hu2008}
M. Hu, P. Keblinski, J.-S. Wang, and N. Raravikar,
\newblock {\em Interfacial thermal conductance between silicon and a vertical carbon
  nanotube},
\newblock J. Appl. Phys. {\bf 104}, 083503 (2008).

\bibitem{Dhar-Roy-JSP-2008}
A.~Dhar and D.~Roy,
\newblock {\em Heat transport in harmonic lattices},
\newblock J. Stat. Phys. {\bf 125}, 801 (2008).


\bibitem{Hu2009}
Jiuning Hu, Xiulin Ruan, and Yong~P. Chen,
\newblock {\em Thermal conductivity and thermal rectification in graphene
  nanoribbons: A molecular dynamics study},
\newblock Nano Lett. {\bf 9}, 2730 (2009).

\bibitem{Guo2010}
Jing Guo, Bin Wen, Roderick Melnik, Shan Yao, and Tingju Li,
\newblock {\em Geometry and temperature dependent thermal conductivity of diamond
  nanowires: A non-equilibrium molecular dynamics study},
\newblock Physica E {\bf   43}, 155 (2010).

\bibitem{Hu2011}
Lin Hu, Tapan Desai, and Pawel Keblinski,
\newblock {\em Determination of interfacial thermal resistance at the nanoscale},
\newblock Physical Review B {\bf 83}, 195423 (2011).

\bibitem{Manikandan2011}
Paranjothy Manikandan, Jeffrey~A. Carter, Dana~D. Dlott, and William~L. Hase,
\newblock {\em Effect of carbon chain length on the dynamics of heat transfer at a
  gold/hydrocarbon interface: Comparison of simulation with experiment},
\newblock J. Phys. Chem. C {\bf 115}, 9622 (2011).

\bibitem{Hsu2007}
W.-D. Hsu, S.~Tepavcevic, L.~Hanley, and S. B. Sinnott,
\newblock {\em Mechanistic studies of surface polymerization by ion-assisted
  deposition},
\newblock J. Phys. Chem. C {\bf 111}, 4199 (2007).

\bibitem{Zwanzig2001}
R.~Zwanzig,
\newblock {\em Nonequilibrium Statistical Mechanics},
\newblock Oxford University Press, 2001.

\bibitem{my-SBC-1}
L.~N. Kantorovich,
\newblock {\em Generalized Langevin equation for solids: I Rigorous derivation and
  main properties},
\newblock Phys. Rev. B {\bf 78}, 094304 (2008).

\bibitem{Lorenzo-GLE-2014}
L.~Stella, C.~D. Lorenz, and L.~Kantorovich,
\newblock {\em Generalized Langevin equation: An efficient approach to
  nonequilibrium molecular dynamics of open systems},
\newblock Phys. Rev. B {\bf 89}, 134303 (2014).

\bibitem{Herve-GLE-2015}
H.~Ness, L.~Stella, C.~D. Lorenz, and L.~Kantorovich,
\newblock {\em Applications of the generalized Langevin equation: Towards a
  realistic description of the baths},
\newblock Phys. Rev. B {\bf 91}, 014301 (2015).

\bibitem{Herve-GLE-2016}
H.~Ness, A.~Genina, L.~Stella, C.~D. Lorenz, and L.~Kantorovich,
\newblock {\em Nonequilibrium processes from generalised Langevin equations:
  realistic nanoscale systems connected to two thermal baths},
\newblock Phys. Rev. B {\bf 93}, 174303 (2016).

\bibitem{FKM-J.Math.Ph.-1965}
G.~W. Ford, M.~Kac, and P.~Mazur,
\newblock {\em Statistical mechanics of assemblies of coupled oscillator},
\newblock J. Math. Phys. {\bf 6}, 504 (1965).

\bibitem{Benguria-Kac-1981}
Rafael Benguria and Mark Kac,
\newblock {\em Quantum Langevin equation},
\newblock Phys. Rev. Lett. {\bf 46}, 1 (1981).

\bibitem{Lindenberg-West-1984}
Katja Lindenberg and Bruce~J. West,
\newblock {\em Statistical properties of quantum systems: The linear oscillator},
\newblock Phys. Rev. A {\bf 30}, 568 (1984).

\bibitem{Cortes-West-Lindenberg-JCP-1985}
E.~Cortes, B.~J. West, and K.~Lindenberg,
\newblock {\em On the generalised Langevin equation: classical and quantum
  mechanical},
\newblock J. Chem. Phys. {\bf 82}, 2708 (1985).

\bibitem{Ford-Kac-JST-1987}
G.~W. Ford and M.~Kac,
\newblock {\em On the quantum Langevin equation},
\newblock J. Stat. Phys. {\bf 46}, 803 (1987).

\bibitem{Ford-Lewis-PRB-1988}
G.~W. Ford, J.~T. Lewis, and R.~F. O'Connell,
\newblock {\em Quantum Langevin equation},
\newblock Phys. Rev. B {\bf 37}, 4419 (1988).

\bibitem{Gardiner-1988}
C.~W. Gardiner,
\newblock {\em Quantum noise and quantum Langevin equations},
\newblock IBM J. Res. Develop. {\bf 32}, 127 (1988).

\bibitem{van_Kampen-1997}
N.~G. van Kampen,
\newblock {\em Derivation of the quantum Langevin equation},
\newblock J. Molec. Liquids {\bf 71}, 97 (1997).

\bibitem{McDowell-JCP-2000}
H.~K. McDowell,
\newblock {\em Quantum generalized Langevin equation: Explicit inclusion of
  nonlinear system dynamics},
\newblock J. Chem. Phys. {\bf 112}, 6971 (2000).

\bibitem{Nieuwenhuizen-PRE-2002}
Th.~M. Nieuwenhuizen and A.~E. Allahverdyan,
\newblock {\em Statistical thermodynamics of quantum Brownian motion: Construction
  of perpetuum mobile of the second kind},
\newblock Phys. Rev. E {\bf 66}, 036102 (2002).

\bibitem{Yan-Xu-review-2005}
Yi~Jing Yan and Rui~Xue Xu,
\newblock {\em Quantum mechanics of dissipative systems},
\newblock Annu. Rev. Phys. Chem. {\bf 56}, 187 (2005).

\bibitem{Gardiner-book_noise-2010}
C.~W. Gardiner and P.~Zoller,
\newblock {\em Quantum noise},
\newblock Springer, 2010.

\bibitem{Ferialdi-Durr-PRA-2015}
L.~Ferialdi and D.~D{\"u}rr,
\newblock {\em Progress towards an effective non-Markovian description of a system
  interacting with a bath},
\newblock Phys. Rev. A {\bf 91}, 042130 (2015).

\bibitem{Segal2002}
D.~Segal and A.~Nitzan,
\newblock {\em Heating in current carrying molecular junctions},
\newblock J. Chem. Phys. {\bf 117}, 3915 (2002).

\bibitem{Segal-Nitzan-Hanggi-2003}
D.~Segal, A.~Nitzan, and P.~Hanggi,
\newblock {\em Thermal conductance through molecular wires},
\newblock J. Chem. Phys. {\bf 119}, 6840 (2003).

\bibitem{Ozpineci-Ciraci-2001}
A.~Ozpineci and S.~Ciraci,
\newblock {\em Quantum effects of thermal conductance through atomic chains},
\newblock Phys. Rev. B {\bf 63}, 125415 (2001).

\bibitem{Wang-Wang-Zeng-2006}
J.-S. Wang, J.~Wang, and N.~Zeng,
\newblock {\em Nonequilibrium greens function approach to mesoscopic thermal
  transport},
\newblock Phys. Rev. B {\bf 74}, 033408 (2006).

\bibitem{Galperin-Nitzan-Ratner-2007}
M.~Galperin, A.~Nitzan, and M.~A. Ratner,
\newblock {\em Heat conduction in molecular transport junctions},
\newblock Phys. Rev. B {\bf 75}, 155312 (2007).

\bibitem{Feynman-Vernon-1963}
R.~P. Feynman and F.~L. Vernon,
\newblock {\em The theory of a general quantum system interacting with a linear
  dissipative system},
\newblock Ann. Phys. {\bf 24}, 118 (1963).

\bibitem{Sebastian-CPL-1981}
K.~L. Sebastian,
\newblock {\em The classical description of scattering from a quantum system},
\newblock Chem. Phys. Lett. {\bf 81}, 14 (1981).

\bibitem{Eckern-JSP-1990}
U.~Eckern, W.~Lehr, A.~Menzel-Dorwarth, F.~Pelzer, and A.~Schmid,
\newblock {\em The quasiclassical langevin equation and its application to the decay
  of a metastable state and to quantum fluctuations},
\newblock J. Stat. Phys. {\bf 59}, 885 (1990).

\bibitem{Kleinert-Shabanov-PLA-1995}
H.~Kleinert and S.V. Shabanov,
\newblock {\em Quantum Langevin equation form forward-backward path integral},
\newblock Phys. Lett. A {\bf 200}, 224 (1995).

\bibitem{Wang-PRL-2007}
J.-S. Wang,
\newblock {\em Quantum thermal transport from classical molecular dynamics},
\newblock Phys. Rev. Lett. {\bf 99}, 160601 (2007).

\bibitem{Dammak-PRL-2009}
H.~Dammak, Y.~Chalopin, M.~Laroche, M.~Hayoun, and J.-J. Greffet,
\newblock {\em Quantum thermal bath for molecular dynamics simulation},
\newblock Phys. Rev. Lett. {\bf 103}, 190601 (2009).

\bibitem{Banerjee-Bag-Banik-Roy-PRE-2002}
D.~Banerjee, B.~C. Bag, S.~K. Banik, and D.~S. Ray,
\newblock {\em Approach to quantum kramers equation and barrier crossing dynamics},
\newblock Phys. Rev. E {\bf 65}, 021109 (2002).

\bibitem{Banerjee-Banik-Bag-Ray-PHE-2002}
D.~Banerjee, S.~K. Banik, B.~C. Bag, and D.~S. Ray,
\newblock {\em Quantum Kramers equation for energy diffusion and barrier crossing
  dynamics in the low-friction regime},
\newblock Phys. Rev. E {\bf 66}, 051105 (2002).

\bibitem{Banik-Bag-Ray-PRE-2002}
S.~K. Banik, B.~C. Bag, and D.~S. Ray,
\newblock {\em Generalized quantum Fokker-Planck, diffusion, and Smoluchowski
  equations with true probability distribution functions},
\newblock Phys. Rev. E {\bf 65}, 051106 (2002).

\bibitem{Banerjee-Bag-Banik-Ray-JCP-2004}
D.~Banerjee, B.~C. Bag, S.~K. Banika, and D.~S. Ray,
\newblock {\em Solution of quantum Langevin equation: Approximations, theoretical
  and numerical aspects},
\newblock J. Chem. Phys. {\bf 120}, 8960 (2004).


\bibitem{Bag-Banik-explanation}
In Refs. \cite{Banerjee-Bag-Banik-Roy-PRE-2002,Banerjee-Banik-Bag-Ray-PHE-2002,Banik-Bag-Ray-PRE-2002,Banerjee-Bag-Banik-Ray-JCP-2004} the Heisenberg operator equations of motion for
the open system are first quantum-averaged using a predefined state of
the whole system. The chosen wavefunction of this state is an Ansatz containing a  product of
an arbitrary state of the open system and a product of coherent states
of each harmonic oscillator of the bath; no justification of why this
particular product of states has been taken is provided. 
Then, they take a quantum-mechanical average of the operator equation of motion  (instead of the 
quantum statistical-mechanical averaging), thereby arriving at a $c$-number Langevin equation.
Finally,  the correlation function 
of the random force is defined using a Gaussian-like Wigner distribution 
for a set of independent shifted harmonic oscillators.
Again, there is no
clear justification of why one should use such a procedure.

\bibitem{Caldeira-Leggett-1983}
A.~O. Caldeira and A.~J. Leggett,
\newblock {\em Path integral approach to quantum Brownian motion},
\newblock Physics A {\bf 121}, 587 (1983).

\bibitem{Zubarev-1}
D.~Zubarev, V.~Morozov, and G.~R\"opke,
\newblock {\em Statistical mechanics of nonequilibrium processes. Vol. 1: Basic
  concepts, kinetic theory},
\newblock Akademie verlag, Berlin, 1996.

\bibitem{CEID-JPCM-2005}
A.~P. Horsfield, D.~R. Bowler, A.~J. Fisher, T.~N. Todorov, and C.~G.
  S{\'a}nchez,
\newblock {\em Correlated electron ion dynamics: the excitation of atomic motion by
  energetic electrons},
\newblock J. Phys.: Condens. Matter {\bf 17}, 4793--4812 (2005).

\bibitem{Gaspard-JCP-1999}
P.~Gaspard and M.~Nagaoka,
\newblock {\em Non-Markovian stochastic Schr\"{o}dinger equation},
\newblock J. Chem. Phys. {\bf 111}, 5676 (1999).

\bibitem{Risken}
H.~Risken,
\newblock {\em The Fokker-Planck equation. Methods of solution and
  applications},
\newblock Springer-Verlag, Berlin, 2nd edition, 1989.

\bibitem{Habershon-ARPC-2013}
S.~Habershon, D.~E. Manolopoulos, T.~E. Markland, and T.~F.~Miller III,
\newblock {\em Ring-polymer molecular dynamics: Quantum effects in chemical dynamics
  from classical trajectories in an extended phase space},
\newblock Annu. Rev. Phys. Chem. {\bf 64}, 387--413 (2013).

\bibitem{Herrero-Ramirez-JPCM-2014}
C.~P. Herrero and R.~Ram{\'i}rez,
\newblock {\em Path-integral simulation of solids},
\newblock J. Phys.: Condens. Matter {\bf 26}, 233201 (2014).

\end{thebibliography}

\end{document}